\documentclass[manuscript, screen, acmsmall, nonacm]{acmart}
\usepackage{subcaption}
\usepackage{hyperref}
\usepackage{booktabs}
\usepackage{pifont}
\usepackage{xcolor}
\definecolor{BrickRed}{RGB}{203,65,84}      
\definecolor{ForestGreen}{RGB}{34,139,34}
\usepackage{colortbl}        
\usepackage{adjustbox}

\usepackage{pdflscape}

\renewcommand\footnotetextcopyrightpermission[1]{} 
\usepackage[skins, breakable]{tcolorbox}
\AtBeginDocument{%
  }

\begin{document}

\title{Generative AI and Creativity: A Systematic Literature Review and Meta-Analysis}

\author{Niklas Holzner}
\email{n.holzner@campus.lmu.de}
\affiliation{
  \institution{LMU Munich}
  \city{Munich}
  \state{Bavaria}
  \country{Germany}
}
\author{Sebastian Maier}
\email{maier.sebastian@campus.lmu.de}
\affiliation{
  \institution{LMU Munich}
  \city{Munich}
  \state{Bavaria}
  \country{Germany}
}
\author{Stefan Feuerriegel}
\email{feuerriegel@lmu.de}
\affiliation{
  \institution{LMU Munich \& Munich Center for Machine Learning (MCML)}
  \city{Munich}
  \state{Bavaria}
  \country{Germany}
}
\begin{abstract}
Generative artificial intelligence (GenAI) is increasingly used to support a wide range of human tasks, yet empirical evidence on its effect on creativity remains scattered. Can GenAI generate ideas that are creative? To what extent can it support humans in generating ideas that are both creative and diverse? In this study, we conduct a meta-analysis to evaluate the effect of GenAI on the performance in creative tasks. For this, we first perform a systematic literature search, based on which we identify $n$ = 28 relevant studies ($m = 8214$ participants) for inclusion in our meta-analysis. We then compute standardized effect sizes based on Hedges' $g$. We compare different outcomes: (i)~how creative GenAI is; (ii)~how creative humans augmented by GenAI are; and (iii)~the diversity of ideas by humans augmented by GenAI. Our results show no significant difference in creative performance between GenAI and humans ($g = -0.05$), while humans collaborating with GenAI significantly outperform those working without assistance ($g = 0.27$). However, GenAI has a significant negative effect on the diversity of ideas for such collaborations between humans and GenAI ($g = -0.86$). We further analyze heterogeneity across different GenAI models (e.g., GPT-3.5, GPT-4), different tasks (e.g., creative writing, ideation, divergent thinking), and different participant populations (e.g., laypeople, business, academia). Overall, our results position GenAI as an augmentative tool that can support, rather than replace, human creativity---particularly in tasks benefiting from ideation support.
\end{abstract}

\maketitle

\section{Introduction}
\label{sec:Introduction}

Generative artificial intelligence (GenAI) refers to a class of machine learning technologies that have the capability to generate new content that resembles human-created output, such as images, text, audio, and videos \cite{Feuerriegel.2024}. GenAI can thus support various human tasks such as writing, software development, composing lyrics, and academic research, often at a performance similar to that of humans \cite{Grimes.2023,Herbold.2023,Cui.2024,Zhou.2024,Ruksakulpiwat.2024,Brynjolfsson.2025}. On top of that, Generative AI has also become a valuable tool in creative industries---spanning graphic design, advertising, fashion, writing, and visual arts \cite{Kapoor.0,Sun.2024}. 

Yet, empirical evidence on the benefits of GenAI for creative performance is scattered, and the theoretical arguments are often inconsistent or even contradictory. On the one side, cognitive research argues that creativity is an inherently human trait \cite{Aru.2025,Sb.332024}. One common issue in practice is that GenAI models often lack the tacit knowledge required for systematic, compositional reasoning such as multi-step problem solving to generate ideas perceived as creative, especially in real-world tasks from businesses. Similarly, GenAI is likely to reproduce ideas seen during training rather than ideating novel ideas \cite{Ismayilzada.10222024}. Hence, simply by means of the training data of GenAI, the generated ideas may also be less diverse than those of humans \cite{Doshi.2024b}. On the other side, there is early evidence suggesting that outputs from GenAI are perceived as being creative \cite{Wang.132024}. For example, several studies find benefits from GenAI in creative tasks, but these findings are typically limited to specific creative tasks (e.g., story writing, ideating business models) \cite{Doshi.2024b,Lee.2024,Zhou.2024,Wadinambiarachchi.2024}. This may suggest that GenAI can generally improve creativity, but it is often unclear how well the results generalize across domains.  

Here, we perform a meta-analysis to evaluate the effect of GenAI on creative performance.\footnote{Code and data are available via our Git at \url{https://github.com/SM2982/Meta-Analysis-LLMs-Creativity.git}.} For this, we focus on different outcomes, namely: (i)~how creative GenAI is; (ii)~how creative humans augmented by GenAI are; and (iii)~the diversity of ideas by humans augmented by GenAI.\footnote{We also considered a fourth outcome, namely, the diversity of ideas generated by humans with GenAI support. Yet, our literature search did not return a sufficient number of empirical studies for this outcome. Hence, we refrain from analyzing this outcome. However, we identify this gap as a promising opportunity for future research, which we elaborate on in the discussion section.} To this end, we analyze the following \textbf{research questions (RQs)}:

\begin{itemize}
\item \textbf{RQ1:} \emph{How creative are ideas generated by GenAI (compared to humans without GenAI support)?}
\item \textbf{RQ2a:} \emph{How creative are ideas generated by humans when supported by GenAI (compared to humans without GenAI support)?}
\item \textbf{RQ2b:} \emph{How diverse are ideas generated by humans when supported by GenAI (compared to humans without GenAI support)?}
\end{itemize}

\noindent
To answer the above research questions, we first perform a systematic literature search and then conduct a meta-analysis. Overall, we retrieved $n= 691$ studies for inclusion, and eventually identified $n = 28$ studies with empirical results ($m=8214$ participants). We then calculated the standardized effect size via Hedges' $g$ and computed a random-effects meta-analysis. We further analyzed heterogeneity across different GenAI models (e.g., GPT-3.5, GPT-4), different tasks (e.g., creative writing, ideation, divergent thinking), and different participant populations (e.g., laypeople, business, and academia) to provide a general but differentiated understanding of the effect of GenAI on creativity. 

\section{Methods}
\label{sec:Methods}

In this section, we describe our data collection to identify relevant studies and statistical analysis. 

\subsection{Search Strategy}
    
\textbf{Search databases:} We followed the PRISMA 2020 framework \cite{Page.2021} for systematic literature reviews. To identify relevant studies, we searched the following databases: (i)~Web of Science, (ii)~SSRN, and (iii)~arXiv. Our search includes non-peer-reviewed studies to reflect the rapidly evolving nature of GenAI research and to capture recent advances. Our search was limited to publications in the English language from the past five years, which is loosely aligned with the emergence of foundational models such as GPT and BERT. The knowledge cutoff of our search was May 2, 2025.

\textbf{Search query:} Our search query was intentionally broad to include various ways to relate to GenAI technology and creativity. Overall, our search query was inspired by Schemmer et al.\,(2022) \cite{Schemmer.2022}, which we adapted to our research question, namely, creativity:

  \begin{tcolorbox}[
    enhanced,
    breakable,
    center upper, 
    colback=gray!10,
    colframe=gray!60!black,
    boxrule=0.8pt,
    arc=4pt,
    outer arc=4pt,
    drop shadow={black!50!white,opacity=0.3},
    width=\textwidth,
    fontupper=\footnotesize,
    title=\textbf{Search query}
  ]
    \begin{verbatim}
    TITLE("creativity" OR "creative" OR "ideation" OR "idea")
    AND
    ("AI" OR "Artificial Intelligence" OR "LLM" OR "Large Language Models")
    \end{verbatim}
  \end{tcolorbox}

\begin{figure}[ht]
  \centering
  \includegraphics[width=.8\linewidth]{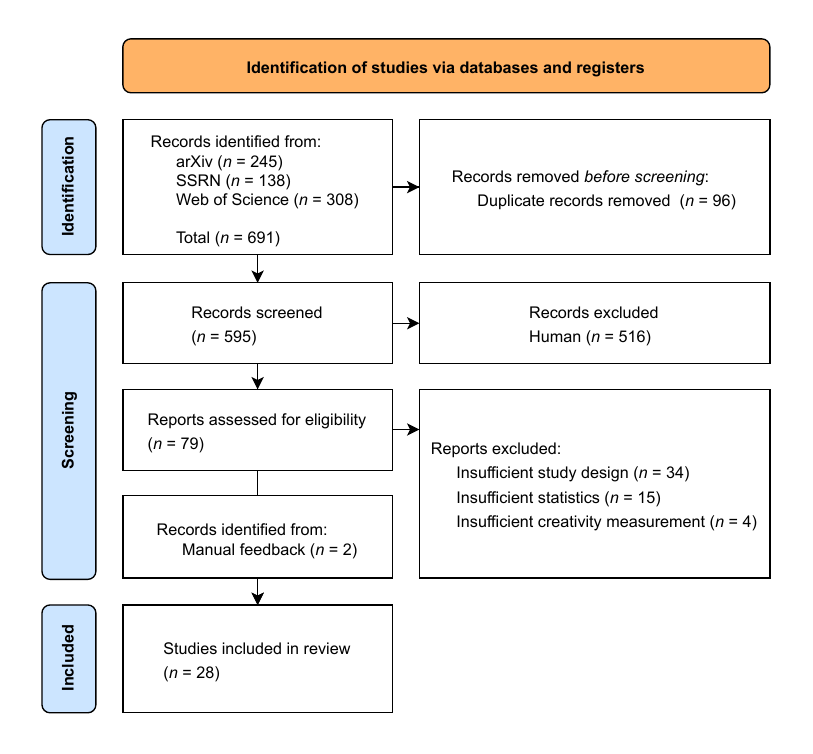}
  \vspace{-0.8cm}
  \caption{\textbf{PRISMA flow chart.}}
  \Description{PRISMA Flow Chart of the Literature Research.}
  \label{fig:PRIMSAFlowchart}
\end{figure}

\textbf{Inclusion/exclusion:} The process for inclusion/exclusion in our systematic literature review is shown in Figure~\ref{fig:PRIMSAFlowchart} (based on the format of the PRISMA 2020 Flowchart \cite{Page.2021b}). Literature screening, eligibility checks, and final inclusion were performed by one person (the first author). In the identification phase, $n = 691$ publications were identified by our search query, out of which $n = 96$ duplicate publications were removed manually. Out of the remaining $n = 595$ publications, both the title and abstract were screened for inclusion in the assessment of eligibility. Here, $n = 516$ publications were excluded due to a lack of fit (e.g., a focus on legal studies). 

Subsequently, $n = 79$ records were assessed for eligibility. Studies were eligible if they:
\begin{enumerate}
  \item The study design was aimed at comparing (a)~the creativity performance of humans versus GenAI or (b)~the creative performance of humans with vs. without GenAI support. The study further followed a between-subject experiment design, which is crucial to make rigorous statistical comparisons.   
  \item The study had to report sufficient statistics (e.g., the group means and standard deviations or equivalent statistics) that allow for computation of standardized effect size Hedges' $g$. In cases where such statistics were not directly reported, we examined whether the publication was accompanied by raw data, either in supplementary materials or associated repositories, that would enable us to reconstruct the necessary statistics. However, if these data were not accompanied by clearly documented analysis scripts and/or the reconstruction would have required substantial interpretation or rewriting of the original code, the study was excluded based on insufficient raw data.
  \item The study had to measure creative performance using an established measurement dimension (e.g., novelty \cite{SusannahB.F.Paletz.2008}, originality \cite{MarkA.Runco.2012}, diversity \cite{Parjanen.2012,SonicRim.2001}). 
\end{enumerate}
In this stage, publications were excluded due to (i)~insufficient study design ($n = 34$) (ii)~insufficient statistics ($n = 15$), and (iii)~insufficient creativity measurement ($n = 4$). Cases with ambiguity were resolved through discussion until consensus was reached (based on Author \#1 together with Author \#2 and Author \#3).

We contacted the corresponding authors from 15 publications via e-mail to address two of the above reasons for exclusion, namely, (ii) insufficient statistics and (iv) insufficient data reporting. If no response was received after the initial contact, a follow-up e-mail was sent to maximize the inclusion of eligible studies. As a result, we obtained sufficient additional information for inclusion from four studies ($n = 4$). All studies that did not elicit a response were excluded.

Given the novel scope of the research question and the evolving terminology in the field, we recognized the possibility that the search strategy might overlook relevant studies. To address this, the authors manually included two additional studies that met all eligibility criteria but were not captured by the original database query.

In total, we identified $n = 28$ publications that met the eligibility criteria for inclusion in our meta-analysis. Several of these publications reported multiple experiments or included multiple outcome measures related to creativity and diversity. As a result, the 28 studies correspond to a total of 127 effect size estimates.

\subsection{Data Collection}

The $n$ = 28 studies that met the inclusion criteria were recorded in our database. Our database capturing all relevant study features and outcomes is deposited in our GitHub. 

An overview of the extracted dimensions is summarized in Table~\ref{tab:variables}, which serves as the basis for exploring heterogeneity in creative performance across various study and task characteristics. Specifically, we extracted study-level data of metadata (title, author, abstract, publication date, source). We further provide details about the experimental design  (GenAI type, GenAI model, task type, creativity measurement, evaluator), participant characteristics (professional domain, recruitment source), and outcome statistics (means, $SD$, $SE$, $m_\mathrm{total}$, $m_\mathrm{control}$, $m_\mathrm{treatment}$, $F$-value, standardized $\beta$, and $SE-\beta$). We discuss the dimensions in the following.

The \textbf{task type} reflects the notion that different creative tasks tap into distinct aspects of creativity (and thus different cognitive processes), such as compositional skills, imaginative reasoning, or divergent thinking (i.e., the ability to generate multiple, varied ideas in response to an open-ended prompt). The studies in our analysis employ a variety of creative tasks, each targeting different dimensions of creative cognition. For example, the so-called \emph{alternate uses task} (AUT) \cite{Guilford.1978} asks participants to generate unconventional uses for everyday objects (e.g., ``a brick''), which assesses originality. The so-called \emph{consequences task (CT)} \cite{Wilson.1953} involves imagining outcomes of improbable scenarios (e.g., ``What would happen if people could fly?''), which taps into imaginative thinking and hypothetical reasoning. The so-called \emph{divergent association task (DAT)} \cite{JayA.Olson.2021} requires listing unrelated words (e.g., ``apple'', ``justice'', ``galaxy''). \emph{Forward flow} (FF) measures \cite{Gray.2019} the conceptual distance between sequential thoughts in open-ended responses, reflecting the natural flow of creative ideation. Finally, other studies rely upon creative writing tasks such as composing a short story or poetry to challenge narrative creativity, but also business ideation tasks (e.g., coming up with a new business model).

Prior work has suggested that creative performance comprises multiple dimensions \cite{MarkA.Runco.2012}. Hence \textbf{creativity measurements} can be obtained in subjective (e.g., human ratings) or objective (e.g., via metrics from natural language processing such as semantic distance or cosine similarity \cite{Feuerriegel.2025}) ways. Subjective assessments commonly rely on scales such as perceived creativity, originality, or novelty---terms that show some overlap conceptually but are occasionally operationalized as distinct constructs \cite{JonathanA.Plucker.2004}. For example, \emph{creativity} is often perceived as a combination of novelty and effectiveness, \emph{originality} refers to how uncommon or unique an idea is, and \emph{novelty} emphasizes newness or unfamiliarity. These dimensions may be interpreted differently by human raters, depending on context and individual experience. Hence we also coded the identity of the \textbf{evaluator}. For example, self-assessments may introduce bias due to over- or underestimation of one's creative performance, in contrast to expert ratings. Rule-based or AI-based evaluations offer standardization and scalability but may miss nuanced judgment or (tacit) domain knowledge. 

In cases where studies reported multiple measurements per experiment, we prioritized evaluations of creative performance by experts over self-assessments. When multiple experiments were reported within a single study, each experiment was included separately, with one outcome measurement per experiment. Creative performance was encoded using the most conceptually appropriate metric available---ideally an overall creativity score (if available) or, alternatively, measures of originality or novelty before using any other measurement.\footnote{To examine potential heterogeneity across different conceptualizations of creative performance, we conducted exploratory subgroup analyses based on the specific constructs used---namely, creativity, originality, and novelty. However, these analyses revealed no substantial heterogeneity. This may be due to the high conceptual overlap among the constructs of creativity, originality, and novelty, which are often used interchangeably in both academic and applied contexts. We thus omitted the analysis for space but the reproducibility code is available for interested readers in our repository.}

\begin{table}[H]
  \centering
  \caption{\textbf{Extracted dimensions.}\label{tab:variables}}
  \vspace{-0.4cm}
  \footnotesize
  \begin{tabular}{@{\extracolsep{\fill}} l  p{0.7\linewidth} }
    \toprule
    \textbf{Dimensions} & \textbf{Values} \\
    \midrule
    GenAI type & text-to-text (T2T), text-to-image (T2I) \\
    \midrule
    GenAI model & GPT-4o, GPT-4, GPT-4all, GPT-3.5-turbo, GPT-3.5, GPT-3, Claude, SparkDesk, Qwen, Dou\,Bao, alpaca, bing, dolly, koala, oa, stablelm, vicuna, not disclosed (n.d.); \emph{optional:} multiple models \\
    \midrule
    Participants & academia, business, laypeople, not disclosed \\
    \midrule
    Task type & alternate uses task (AUT), consequences task (CT), divergent associations task (DAT), forward flow (FF), creative writing, creative problem solving, creative thinking, divergent thinking, ideation product, ideation item usage, ideation research proposal, ideation business concepts, ideation other\\
    \midrule
    Recruitment source & university, Prolific, Mturk, public, company \\
    \midrule
    Creativity measurement & creativity scale, originality scale, novelty scale, semantic distance, cosine similarity, creative problem-solving scale, flexibility score \\
     \midrule
    Evaluator & self-assessed, laypeople, expert, rule-based, AI \\
    \addlinespace
    \bottomrule
  \end{tabular}
\end{table}

\subsection{Statistical Analysis}
\label{sec:StatAnalysis}

\textbf{Hedges' $g$:} For each comparison, we extracted reported Cohen's $d$ or calculated it based on reported statistics \cite{Cohen.2013}. When effect sizes were not directly reported, we converted other statistical measures (e.g., $t$-values, $F$-values, means and standard deviations) to Cohen's $d$ using widely accepted conversion formulas \cite{Borenstein.2021,Dunlap.1996,Peterson.2005,Rosnow.1996}, which are documented in our project repository. To adjust for potential upward bias in small samples, we applied Hedges' $g$ correction \cite{LarryV.Hedges.1981}. We report one effect size per individual experiment across all included studies to ensure statistical independence. We further report 95\% confidence intervals (CIs).

\textbf{Random-effects model:} As we expect large heterogeneity regarding treatment, task, and measurement across studies, we chose to estimate a random-effects model as recommended by Cochrane \cite{Deeks.2019}. Pooled effect sizes were estimated under the random-effects model using the DerSimonian-Laird estimator \cite{DerSimonian.1986} to account for the expected between-study heterogeneity. We calculated 95\% confidence intervals (CIs) to indicate the expected range of true effects across settings. 

\textbf{Variability ($I^2$ statistic):} The variability in effect sizes across studies was quantified using the $I^2$ statistic, Cochran's $Q$ test, and by estimating the between-study variance ($\tau^2$) via Jackson's method \cite{Jackson.2016}. 

\textbf{Heterogeneity analysis:} To further explore potential sources of heterogeneity, we conducted subgroup analyses, such as comparing creativity across participants with different backgrounds (e.g., academic vs. business). As a robustness check, we also performed meta-regression analyses \cite{Viechtbauer.2010b}, which yielded results consistent with the subgroup analyses. For reasons of space, detailed meta-regression results are provided in our GitHub repository.

\textbf{Bias assessment:} Risk of bias was assessed using the Cochrane Risk-of-Bias 2.0 tool \cite{Sterne.2019}, with judgments recorded for selection, performance, detection, and reporting bias. Publication bias was evaluated through Egger's regression test \cite{Egger.1997} and the trim-and-fill procedure \cite{Duval.2000} for comparisons including at least ten studies. These analyses revealed no substantial concerns regarding bias; we thus included full results in our GitHub for brevity.

\textbf{Implementation details:} All analyses were implemented in R (version 4.2.3) using the \texttt{metafor} package (version 4.8-0). Influence diagnostics were conducted via the \texttt{influence()} function from the \texttt{metafor} package, and leave-one-out sensitivity analyses \cite{Viechtbauer.2010} were used to assess the robustness of pooled estimates. Codes are available in our repository for reproducibility. 

\newpage

\begin{landscape}

\begin{table}[H]
  \centering
    \tiny
    \fontsize{5}{5.8}\selectfont
    \setlength{\tabcolsep}{1pt}
    \renewcommand{\arraystretch}{1.0}
    \rowcolors{2}{lightgray!40}{white}  
    \begin{tabular}{l | ll ll ll | p{6cm} | p{2cm} | l | l | p{2cm} | p{0.5cm} | p{0.9cm} | p{2cm} | p{1.5cm}}
\rowcolor{gray!30} 
\hline
ID & HAI ID & & CP ID & & CD ID & & Title & GenAI model & Participants & \#Participants ($m$) & Task type & GenAI type & Recruitment source & Creativity measurement & Evaluator \\
\hline\hline
1 & HAI01 & R12 &&&&& A Confederacy of Models: a Comprehensive Evaluation of LLMs on Creative Writing \cite{GomezRodriguez.10122023} & GPT-4, GPT-3.5, bing, claude12, koala, vicuna, oa, bard, GPT-4all, stablelm, dolly, alpaca & Academia & 5 & creative writing & T2T & University & creativity scale & Expert \\
2 & HAI02 & R6 &&&&& AI Delivers Creative Output but Struggles with Thinking Processes \cite{Zhang.3302025} & GPT-3.5, GPT-4, GPT-4o & Laypeople & 162 & AUT & T2T & Public & novelty scale & Human \\
3 & HAI03 & R1 & CP01 & R6 &&& An empirical investigation of the impact of ChatGPT on creativity \cite{Lee.2024}& GPT-3.5 & Laypeople & 1701 & ideation item usage, ideation product, ideation other & T2T & Mturk/Prolific & creativity scale & Expert \\
4 & HAI04 & R1 &&&&& Artificial Creativity? Evaluating AI Against Human Performance in Creative Interpretation of Visual Stimuli \cite{Grassini.2025}& GPT-4 & Not disclosed & 256 & creative interpretation & T2T & Prolific & creativity scale & Human \\
5 & HAI05 & R1 &&&&& Artificial muses: Generative Artificial Intelligence Chatbots Have Risen to Human-Level Creativity \cite{JenniferHaase.2023}& GPT-3, Copy.Ai, Alpa.ai, Studio, YouChat & Academia & 100 & AUT & T2T & Prolific & creativity scale & Human \\
6 & HAI06 & R1 &&&&& Best humans still outperform artificial intelligence in a creative divergent thinking task \cite{Koivisto.2023} & GPT-3.5, GPT-4, Copy.Ai & Not disclosed & 256 & AUT & T2T & Prolific & creativity scale, semantic distance & Human \\
7 & HAI07 & R1 &&&&& Can LLMs Generate Novel Research Ideas? A Large-Scale Human Study with 100+ NLP Researchers \cite{Si.962024} & claude-3.5-sonnet & Academia & 49 & ideation research ideas & T2T & Public & novelty scale & Expert \\
8 & HAI08 & R2 & CP02 & R1 &&& Creative and Strategic Capabilities of Generative Ai: Evidence from Large-Scale Experiments \cite{NoahBohren.2024} & GPT-4, Bard & Not disclosed & 1250 & creative writing & T2T & Prolific & creativity scale & Human \\
9 & HAI09 & R4 &&&&& Creativity and AI \cite{Charness.2024} & GPT-3.5, GPT-4o & Not disclosed & 80 & creative writing, ideation product & T2T & Prolific & creativity scale & Human \\
10 & HAI10 & R1 & CP03 & R3 &&& Establishing the importance of co-creation and self-efficacy in creative collaboration with artificial intelligence \cite{McGuire.2024}& n.d. & Academia & 96 & creative writing & T2T & Prolific & creativity scale & Expert \\
11 & HAI11 & R1 &&& CD03 & R1 & Evaluating Creative Short Story Generation in Humans and Large Language Models \cite{Ismayilzada.1142024} & 60 models & Academia & 59 & creative writing & T2T & Prolific & creativity scale & Expert \\
12 &&& CP04 & R1 &&& Generative AI Adoption in Human Creative Tasks: Experimental Evidence \cite{Zou.2025} & GPT-3.5 & Academia & 246 & ideation other & T2T & University & creativity scale & Human \\
13 &&& CP05 & R2 & CD01 & R2 & Generative artificial intelligence enhances creativity but reduces the diversity of novel content \cite{Doshi.2024b} & GPT-4 & Not disclosed & 293 & creative writing & T2T & Prolific & novelty scale, cosine similarity & Expert / Rule-based \\
14 & HAI12 & R1 &&&&& How AI Ideas Affect the Creativity, Diversity, and Evolution of Human Ideas: Evidence From a Large, Dynamic Experiment \cite{Ashkinaze.1242024}& GPT 3.5 & Not disclosed & 844 & AUT & T2T & Public & originality score & Rule-based \\
15 & HAI13 & R1 &&&&& How AI Outperforms Humans at Creative Idea Generation \cite{Castelo.2024} & GPT-4 & Business & 10 & ideations of new products & T2T & Prolific & creativity scale & Human \\
16 &&& CP06 & R1 & CD02 & R1 & How Experience Moderates the Impact of Generative AI Ideas on the Research Process \cite{Doshi.2024}& GPT-4 & Academia & 310 & ideation research proposal & T2T & University & novelty scale, cosine similarity & Self-assessed \\
17 &&& CP07 & R1 &&& If ChatGPT can do it, where is my creativity? generative AI boosts performance but diminishes experience in creative writing \cite{Mei.2025} & GPT-4o, GPT-4-turbo & Academia & 266 & creative writing & T2T & Prolific & flexibility score & Human \\
18 &&& CP08 & R1 &&& Interactions with generative AI chatbots: unveiling dialogic dynamics, students' perceptions, and practical competencies in creative problem-solving \cite{Song.2025} & Dou Bao & Academia & 80 & creative problem solving & T2T & University & CPS scale & Rule-based \\
19 & HAI14 & R1 & CP09 & R1 & CD04 & R2 & Large Language Model in Creative Work: The Role of Collaboration Modality and User Expertise \cite{Chen.2024} & GPT-4 & Business & 355 & ideation business concepts & T2T & Prolific & creativity scale & Human \\
20 &&& CP10 & R1 && & Large Language Model in Ideation for Product Innovation: An Exploratory Comparative Study \cite{Zheng.2024} & GPT-3.5-turbo & Laypeople & 90 & ideation product & T2T & Mturk & novelty scale & Human \\
21 & HAI15 & R52 &&&&& Large Language Models show both individual and collective creativity comparable to humans \cite{Sun.1242024} & GPT-3.5, GPT-4, Claude, Qwen, SparkDesk & Academia & 467 & Divergent Thinking, problem solving, creative writing & T2T & University & novelty scale, originality scale & Human \\
22 & HAI16 & R1 &&&&& One Does Not Simply Meme Alone: Evaluating Co-Creativity Between LLMs and Humans in the Generation of Humor, 1082–1092 \cite{Wu.2025}& GPT-4o & Academia & 562 & ideation meme content & T2T & Prolific & creativity scale & Human \\
23 & HAI17 & R1 & CP11 & R1 &&& Revolution or inflated expectations? Exploring the impact of generative AI on ideation in a practical sustainability context \cite{AnjaEisenreich.2024} & GPT-4 & Business & 56 & ideation other & T2T & Company & novelty scale & Expert \\
24 &&& CP12 & R2 &&& The Crowdless Future? How Generative AI Is Shaping the Future of Human Crowdsourcing \cite{Boussioux.2024} & GPT-4 & Business & 125 & ideation business concepts & T2T & Public & novelty scale & Expert \\
25 & HAI18 & R5 &&&&& The current state of artificial intelligence generative language models is more creative than humans on divergent thinking tasks \cite{Hubert.2024} & GPT-4 & Academia & 151 & AUT / CT / DAT & T2T & Prolific & semantic distance & AI \\
26 & HAI20 & R2 &&&&& The Language of Creativity: Evidence from Humans and Large Language Models \cite{Orwig.2024}& GPT-3 & Not disclosed & 50 & creative writing & T2T & Prolific & creativity scale & Human \\
27 & HAI21 & R3 &&&&& We're Different, We're the Same: Creative Homogeneity Across LLMs \cite{Wenger.1312025} & 22 models & Not disclosed & 102 & AUT / FF / DAT & T2T & Prolific & semantic distance & Rule-based \\
28 & HAI22 & R2 &&&&& Writing, creativity, and artificial intelligence. ChatGPT in the university context \cite{deVicenteYagueJaraMarinezOlivia.2023} & 20 models & Academia & 193 & ideation item usage, creative thinking & T2T & University & originality score & Rule-based \\
\hline
\end{tabular}
  \caption{\textbf{Overview of studies included by the structured literature review on GenAI and creativity. The columns HAI, CP, and CD note the identification number of each study in the different meta-analyses across the different outcomes (HAI: creative performance of human vs GenAI; CP: creative performance in human-GenAI collaboration; CD: creative diversity in human-GenAI collaboration). The ``R $X$'' value provided for each study reports how many observations from each study were included in the corresponding meta-analysis (e.g., R3 means that 3 individual studies from the paper were included in the meta-analysis as separate observations).}}
  \label{tab:meta_analysis_literature}
\end{table}

\end{landscape}

\newpage

\section{Results}
\label{sec:results}

We first summarize key characteristics of the included studies, and afterward, we answer our research questions. Note that all analyses are reported at the effect size level (i.e., 127 observations with effect sizes for 28 studies).

\subsection{Descriptive Summary}
\label{sec:descriptive_summary}

\textbf{Study settings:} The studies vary in which outcomes are analyzed. The majority of studies ($n = 21$) measure creative performance between humans versus GenAI (with $m= 4582$ human participants), which is later relevant for \textbf{RQ1}. In contrast, $n = 12$ studies focus on creative performance between humans and human-GenAI collaboration ($m= 2798$; \textbf{RQ2a}), and $n = 4$ studies focus on creative diversity between humans and human-GenAI collaboration ($m= 1017$; \textbf{RQ2b}).

\textbf{GenAI models:} The capabilities of the GenAI model also determine how well it excels with creative tasks. In the included studies, \emph{GPT-4} is the GenAI model that is used most frequently (37 of 127; 29.6\%), followed by \emph{GPT-3.5} (25, 20.0\%). Other GenAI models appear in fewer than 11\% of studies, which, on the one hand, reflects the state-of-the-art performance of GPT-4 in many tasks, but, on the other hand, implies that the findings may not generalize across all GenAI models but are subject to the specific choice (see our discussion in Section~\ref{sec:limitations}).

\textbf{Task type:} The majority of experiments involve tasks aimed at testing \emph{creative writing} to assess creativity (48 of 127 tasks; 38\%), followed by \emph{creative problem solving} (25; 20\%) and the alternate uses test (12, 10\%). In contrast, only a few studies investigate business-oriented ideation tasks (fewer than 15\%).

\textbf{Participants:} The choice of the participant sample determines whether findings later generalize only to laypeople or even to people with domain expertise. Here, the primary participant pool stems from an \emph{academic} background, accounting for over two-thirds of all observations (85 of 127; 67\%), whereas \emph{laypeople} (14; 11\%) and business professionals (9; 7\%) appear far less frequently. Of note, 19 studies do not even disclose the background of their participants or include participant pools with mixed backgrounds.

\subsection{RQ1: Comparing the Creative Performance of Human vs. GenAI}
\label{sec:CreativePerformanceComparisonOfHumanAndAI}

In \textbf{RQ1}, we aim to answer: \emph{How creative are ideas generated by GenAI compared to humans without GenAI support}, we first execute the random-effects model on the studies comparing (a)~the creativity performance of humans versus GenAI. Here, out of the 127 observations, only 100 are relevant as they perform such a comparison.

\begin{figure}[H]
  \centering
  \includegraphics[width=\linewidth,
                 height=0.85\textheight,
                 keepaspectratio]{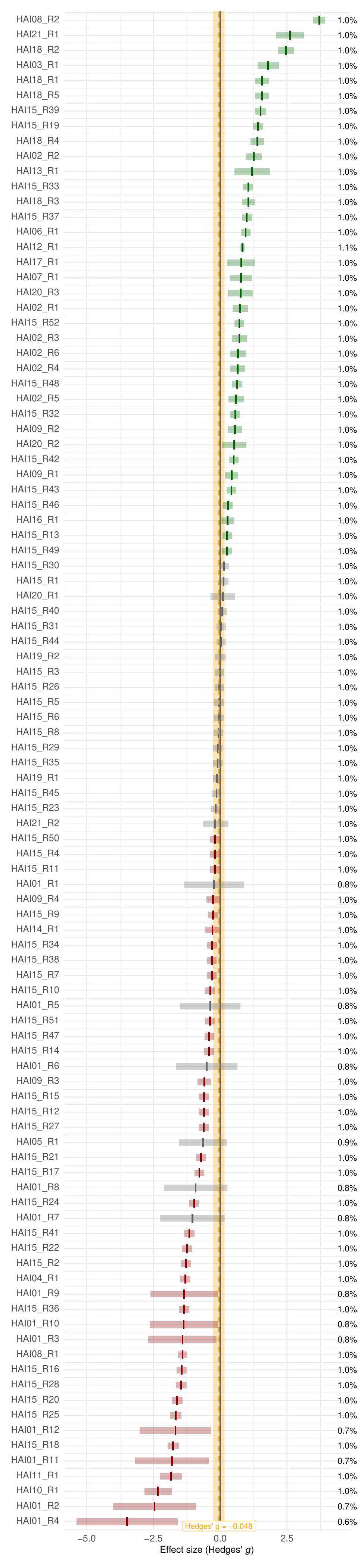}
  \caption{\textbf{Pooled effect comparing the creative performance of humans vs. GenAI (RQ1).} The forest plot summarizes the Hedges' $g$ effect sizes and 95\% confidence intervals for a direct comparison between humans vs. GenAI (treatment: GenAI vs. control: human alone). Out of the 127 observations, 100 observations (participants $m = 4582$) compare differences in creative performance between humans and GenAI, and are thus included in the comparison. Each line is one estimate (the weight is shown at the right). The overall effect size of $g = -0.048$ indicates no statistically significant difference. The vertical line at $g = 0$ corresponds to a null effect; observations to the left favor the human control, and observations to the right favor GenAI. The bars are the estimated effect sizes, and the whiskers are 95\% CIs. The orange dashed line is the mean pooled effect size and the orange shaded area is its 95\% CI.}
  \label{fig:versus_raw_forest}
\end{figure}

\textbf{Pooled effect:} The pooled effect based on our random-effects meta-analysis corresponds to a Hedges'\,$g = -0.048$ (95\% CI:\,[\,$-0.257$, $0.161$]; $p=0.653$). The forest plot is shown in Figure~\ref{fig:versus_raw_forest}. The overall heterogeneity is large ($I^{2}=98.90\%$; $\tau^{2}=1.08$). While the results hint toward a slight disadvantage toward GenAI, the effect sizes show no statistically significant difference between GenAI and human-only conditions across studies.

Each study contributes less than 1.1\% weight, so the finding is not dominated by any single experiment. Nevertheless, we performed a leave-one-out sensitivity analysis. Here, the sequential deletion keeps the pooled estimate between $g=-0.083$ and $g=-0.024$, and with each 95\% CI: still overlapping with zero and the $I^{2}$ staying above 98\%. This shows that the results do not hinge on a single study but corroborate our general finding that there is no or only a negligible difference in the creative performance of humans vs. GenAI. 

\newpage

\subsubsection{Heterogeneity Analysis}
\label{sec:CreativePerformanceComparisonOfHumanAndAI_Moderator}

\begin{figure}[ht]
  \centering
  \begin{subfigure}[t]{0.33\linewidth}
    \centering
    \includegraphics[width=\linewidth]{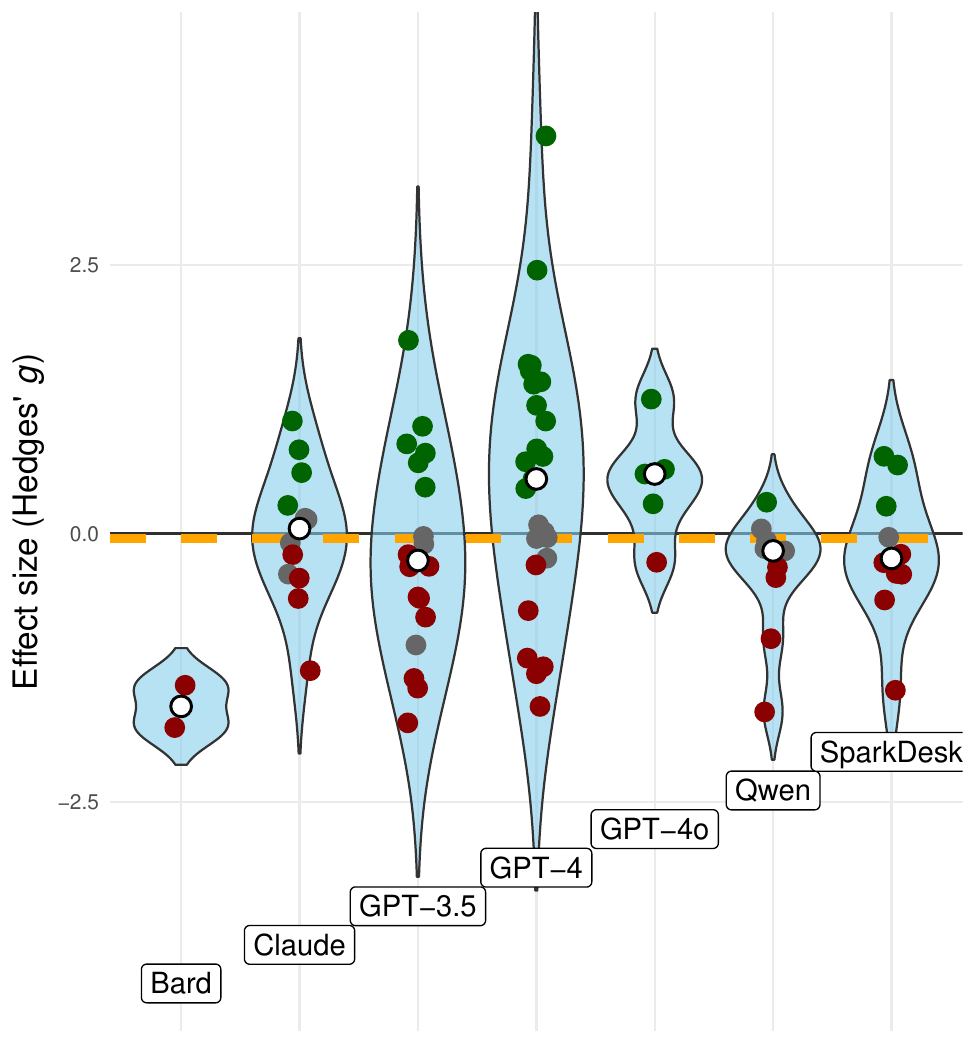}
    \caption{GenAI model.}
    \label{fig:versus_violin_genai_model}
  \end{subfigure}\hfill
  \begin{subfigure}[t]{0.33\linewidth}
    \centering
    \includegraphics[width=\linewidth]{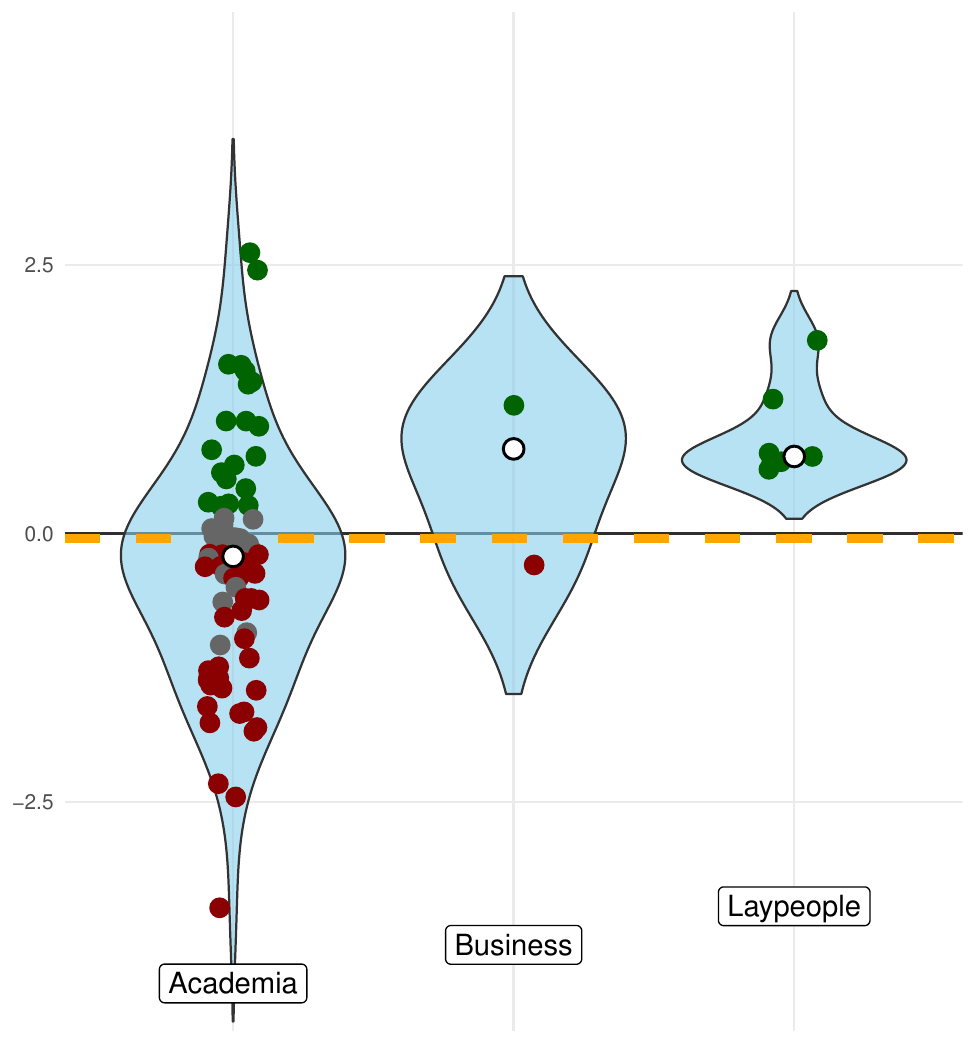}
    \caption{Participant type.}
    \label{fig:versus_violin_participants}
  \end{subfigure}\hfill
  \begin{subfigure}[t]{0.33\linewidth}
    \centering
    \includegraphics[width=\linewidth]{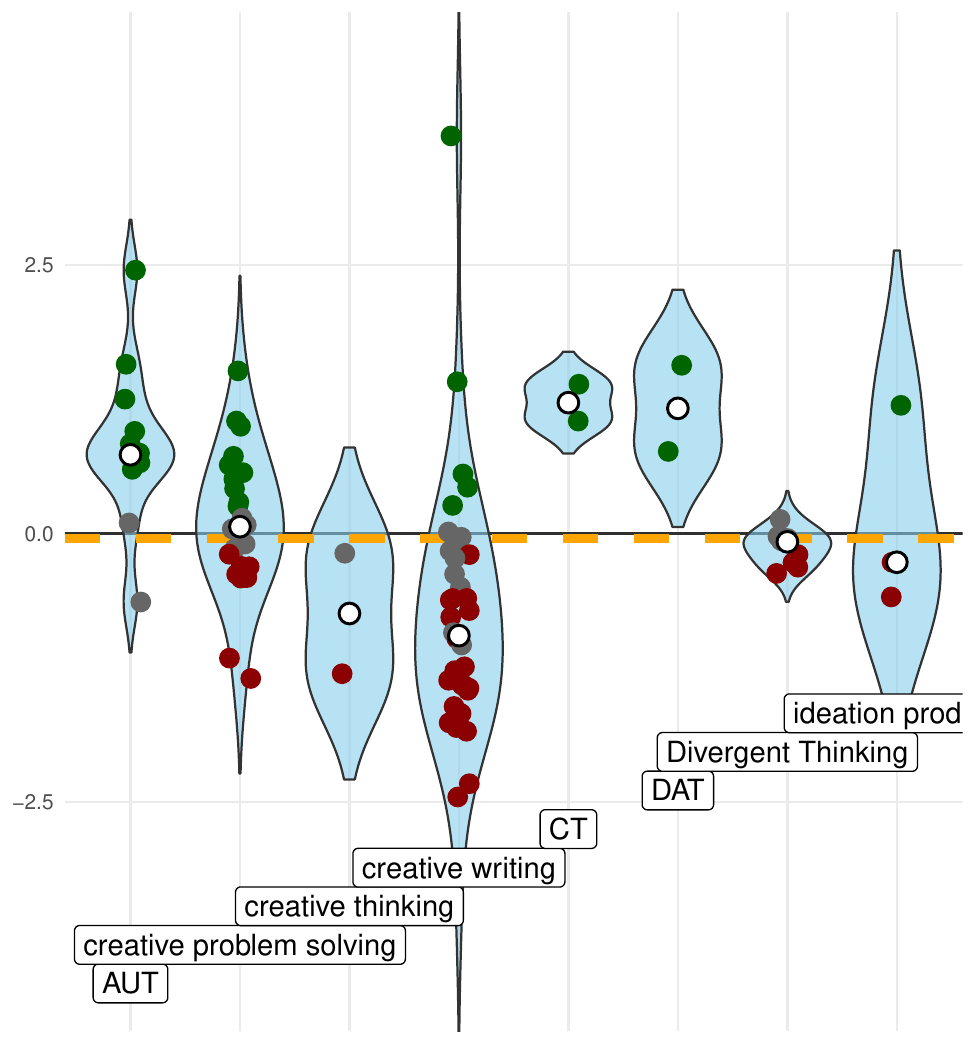}
    \caption{Task type.}
    \label{fig:versus_violin_task_type}
  \end{subfigure}
  \caption{\textbf{Heterogeneity analysis for RQ1 (creative performance of humans vs. GenAI).} Violin plots show the distribution of observation-level Hedges' $g$ for the comparison in creative performance between human vs. GenAI conditions. The comparison is stratified by \textbf{(a)} GenAI model, \textbf{(b)}~participant background, and \textbf{(c)}~ task type. Subgroup analyses are reported only for categories with a sufficient number of observations to support meaningful comparisons. The widths reflect the density of effect sizes; the dashed line corresponds to $g=-0.048$ with no overall difference.}
  \label{fig:RQ1_heterogeneity}
\end{figure}

\textbf{GenAI model (Figure~\ref{fig:RQ1_heterogeneity}a):} As expected, a key determinant for creative performance is the choice of the GenAI model. The model \emph{GPT-4} performs the strongest, with distribution centered slightly above zero. A meta-regression confirms the positive effect for \emph{GPT-4} ($g = 0.499$; 95\% CI: $[0.132, 0.865]$; $p = 0.008$). For comparison, the effect for other models often overlaps with zero and has wide tails, which implies both non-significant coefficients and substantial within-model variability. 

\textbf{Participants (Figure~\ref{fig:RQ1_heterogeneity}b):} \emph{Laypeople} appear to be constantly outperformed by GenAI ($g = 0.918$; 95\% CI: $[0.202, 1.630]$; $p = 0.012$). The distribution of laypeople is right-skewed and centered above zero. While academics show a small but significant pro-human difference ($g = -0.223$; 95\% CI: $[-0.445, -0.001]$; $p = 0.049$), the results for participants with business background remain inconclusive ($g = 0.539$; 95\% CI: $[-0.579, 1.660]$; $p = 0.345$). This suggests that the background of participants (and thus their domain expertise) is a moderator that explains the gap in creative performance between humans and GenAI.

\textbf{Task type  (Figure~\ref{fig:RQ1_heterogeneity}c):} Violin shapes show modes above zero for AUT and CT tasks. The AUT favors GenAI ($g = 0.855$; 95\% CI: $[0.358, 1.350]$; $p < 0.001$), as do CT tasks ($g = 1.220$; 95\% CI: $[0.016, 2.420]$; $p = 0.047$). By contrast, creative-writing tasks favor humans ($g = -0.717$; 95\% CI: $[-1.010, -0.424]$; $p < 0.001$), a pronounced left-shift for creative writing. Other categories (e.g., creative problem solving, divergent thinking) are non-significant, highlighting task-specific effects. The violin plot has a broad overlap for the remaining categories, echoing the task-specific moderator coefficients.

\subsubsection{Robustness}

\textbf{Funnel plot \& Egger's test:} Egger's mixed-effects regression detects significant funnel asymmetry ($z=-3.363$; $p=0.001$), indicating the presence of small-study or publication bias. The \textbf{trim-and-fill adjustment} therefore imputes 24 missing studies on the right of the funnel and shifts the random-effects estimate to ($g=0.364$, 95\% CI:\,[\,$0.131$; $0.596$]; $p=0.002$), implying that bias-correction would reverse the direction of the overall effect in favor of GenAI. 

\textbf{Influence diagnostics:} All influence metrics (studentized residuals, DFFITS, Cook's $D$, hat values, cov.r) remain well below critical cut-offs, except for one potential outlier; removing this study lowers $\tau^{2}$ only modestly (to $0.93$) and leaves the pooled effect virtually unchanged, confirming that no single study drives the results.

\subsection{RQ2a: Benefit of Human-GenAI Collaboration for Creative Performance}
\label{sec:CreativePerformance}

In \textbf{RQ2a}, we answer: \emph{How creative are ideas generated by humans when supported by GenAI (compared to humans without GenAI support)?} Here, we restrict our analysis to $21$ observations where studies report effect sizes aimed at understanding the effect of human-GenAI collaboration.  

\begin{figure}[H]
  \centering
  \includegraphics[width=.5\linewidth,
                 height=0.45\textheight,
                 keepaspectratio]{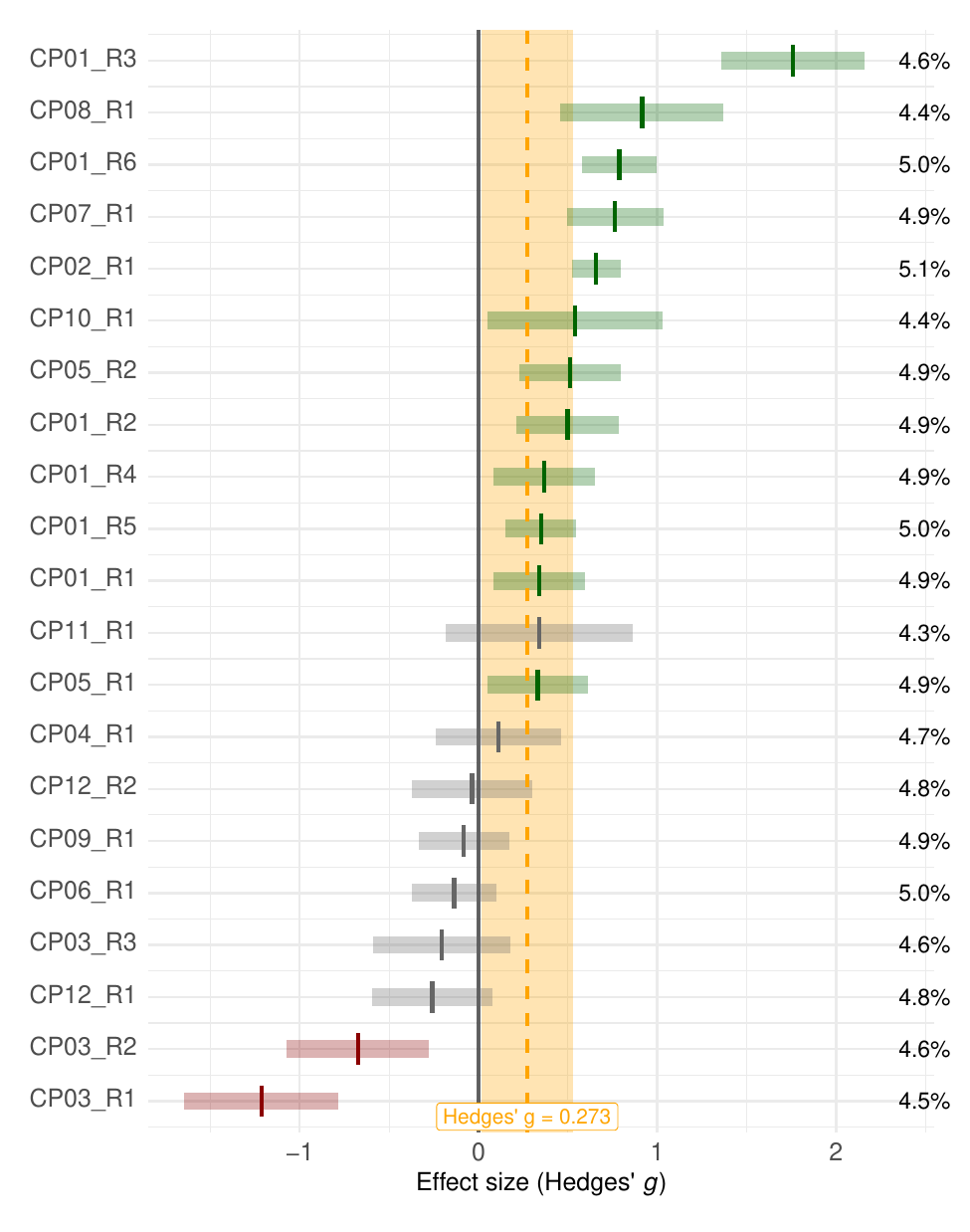}
  \caption{\textbf{Pooled effect of the benefit from Human-GenAI collaboration on creative performance (RQ2a).} The forest plot summarizes the Hedges' $g$ effect sizes and 95\% confidence intervals (treatment: human-GenAI collaboration versus control: human alone). Out of the 127 observations, $n = 21$ observations (participants $m = 2798$) quantify differences in creative performance between humans and human-GenAI collaboration. Each line is one estimate (the weight is shown at the right). The overall effect size of $g = 0.273$ indicates a modest performance gain from GenAI assistance. The vertical line at $g = 0$ corresponds to a null effect; points to the right favor the GenAI-assisted collaboration. The bars are the estimated effect sizes, and the whiskers are the 95\% CIs. The orange dashed line is the mean pooled effect size and the orange shaded area is its 95\% CI.}
  \label{fig:RQ2a_pooled_effect}
\end{figure}

\textbf{Pooled effect:} The pooled effect based on our random-effects meta-analysis amounts to Hedges'\,$g = 0.273$ (95\% CI:\,[0.018, 0.528]; $p = 0.036$). The forest plot is shown in Figure~\ref{fig:RQ2a_pooled_effect}. Again, heterogeneity is substantial ($Q_{20}=232.17$; $p<0.001$; $I^{2}=94.11\%$; $\tau^{2}=0.3261$). This result indicates a small but statistically significant positive effect of GenAI support on the creativity of human-generated ideas, meaning that humans augmented with GenAI are more creative than humans without. 

No single study contributes more than 5.1\% weight, indicating that the positive performance advantage is not driven by any single experiment. Nevertheless, we performed a leave-one-out sensitivity analysis. Sequential deletion keeps the pooled effect between $g = 0.204$ and $g = 0.343$. Across these iterations of the leave-one-out sensitivity analysis, the lower bound of the 95\% CI ranged from $-0.0179$ to $0.1198$, meaning that omitting certain studies can yield a CI that just crosses zero. The $I^{2}$ remained above 91\%, confirming that the positive performance effect is stable yet accompanied by persistent heterogeneity.

\subsubsection{Heterogeneity Analysis}

\begin{figure}[ht]
  \centering
  \begin{subfigure}[t]{0.33\linewidth}
    \centering
    \includegraphics[width=\linewidth]{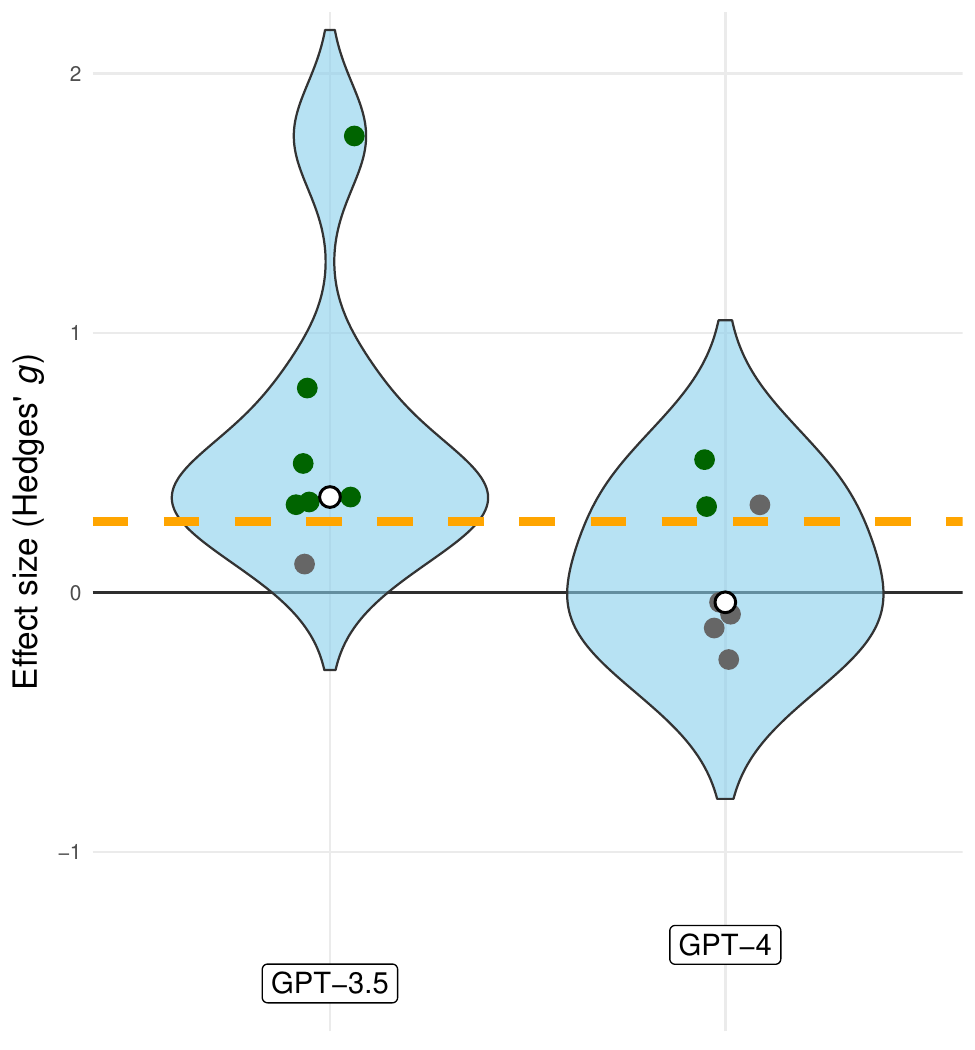}
    \caption{GenAI model.}
    \label{fig:perf_violin_genai_model}
  \end{subfigure}\hfill
  \begin{subfigure}[t]{0.33\linewidth}
    \centering
    \includegraphics[width=\linewidth]{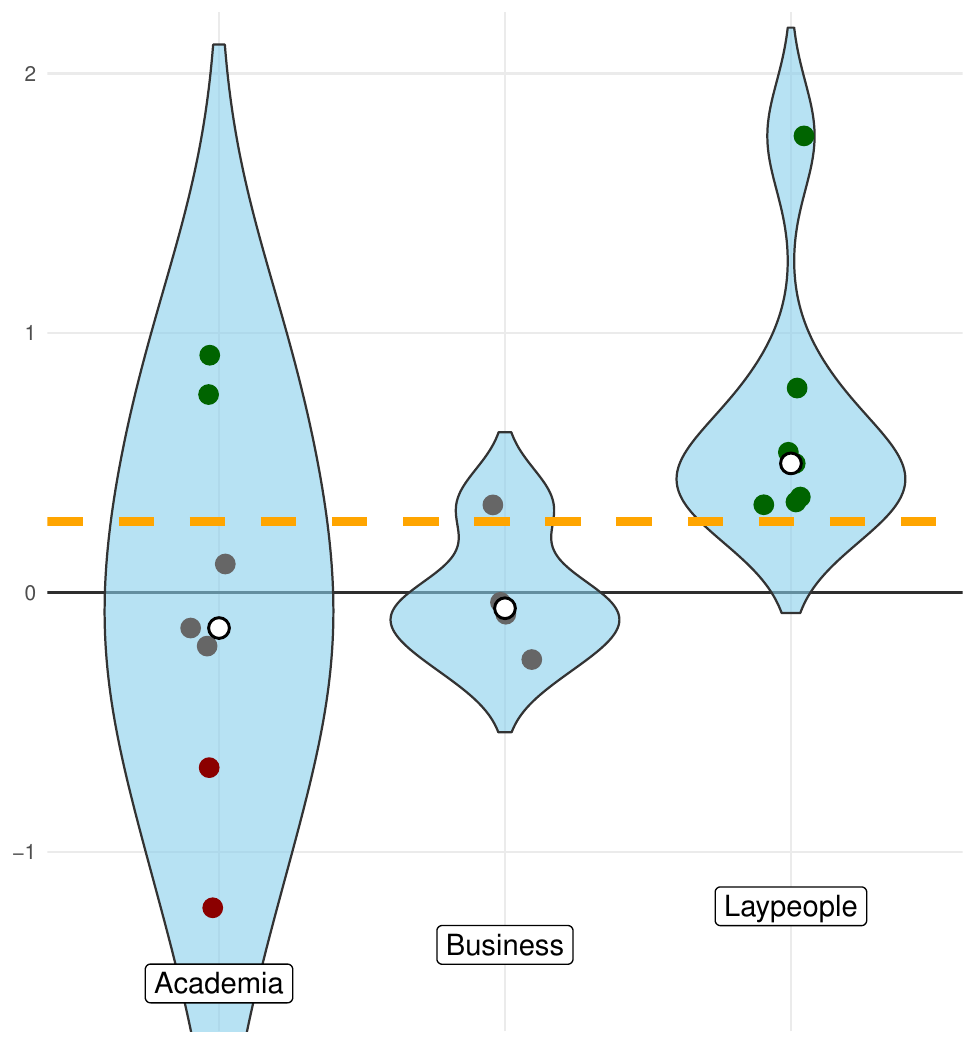}
    \caption{Participant type.}
    \label{fig:perf_violin_participants}
  \end{subfigure}\hfill
  \begin{subfigure}[t]{0.33\linewidth}
    \centering
    \includegraphics[width=\linewidth]{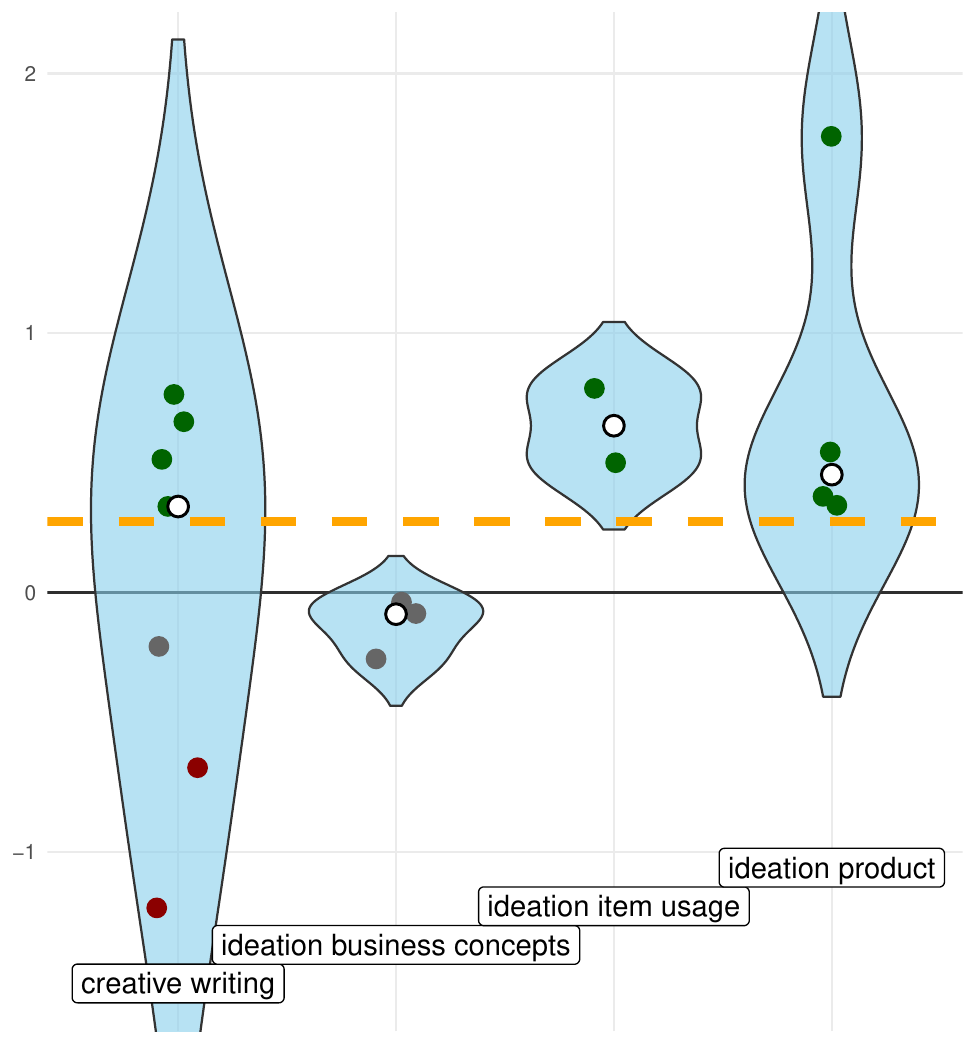}
    \caption{Task type.}
    \label{fig:perf_violin_task_type}
  \end{subfigure}
  \caption{\textbf{Heterogeneity analysis for RQ2a (creative performance of humans+GenAI vs. humans only).} Violin plots show the distribution of observation-level Hedges' $g$ for the benefit in creative performance of human-GenAI collaboration over a human-only condition. The comparison is stratified by \textbf{(a)} GenAI model, \textbf{(b)}~participant background, and \textbf{(c)}~ task type. Subgroup analyses are reported only for categories with a sufficient number of observations to support meaningful comparisons. The widths reflect the density of effect sizes; the dashed line corresponds to $g=0$ with no overall difference.}
  \label{fig:RQ2a_heterogenetiy}
\end{figure}

\textbf{GenAI model (Figure~\ref{fig:RQ2a_heterogenetiy}a):} The plot shows a higher median and thicker right tail for \emph{GPT-3.5} relative to \emph{GPT-4}, which is also confirmed in a meta-analytic regression ($g = 0.587$; 95\% CI: $[0.280, 0.893]$; $p < 0.001$). Interestingly, GPT-3.5 contributes larger performance gains, whereas \emph{GPT-4} shows no detectable deviation from the overall mean ($g = 0.089$; 95\% CI: $[-0.223, 0.401]$; $p = 0.577$).

\textbf{Participants (Figure~\ref{fig:RQ2a_heterogenetiy}b):} The distribution of effect sizes for \emph{laypeople} is right-skewed and centered above zero, point to a benefit from collaboration ($g = 0.654$; 95\% CI: $[0.237, 1.070]$; $p = 0.002$). The distribution of the effect sizes for \emph{academia} ($g = -0.057$; 95\% CI: $[-0.480, 0.367]$) and \emph{business} ($g = -0.021$; 95\% CI: $[-0.581, 0.540]$) cluster around the grand mean, indicating that participant expertise moderates outcomes asymmetrically. In other words, laypeople seem to benefit from co-creation with GenAI, while domain experts do not. 

\textbf{Task type (Figure~\ref{fig:RQ2a_heterogenetiy}c):} A wide and mostly right-shifted violin is found for \emph{ideation-product} tasks ($g = 0.743$; 95\% CI: $[0.128, 1.360]$; $p = 0.018$) where human–GenAI teams show a better creative performance than the human-only condition. The effect sizes for tasks aimed at \emph{creative writing} overlap with zero ($g = 0.048$; 95\% CI: $[-0.412, 0.508]$). The same pattern is found for the other ideation tasks.

\subsubsection{Robustness}

\textbf{Funnel plot \& Egger's test:} Egger's regression detects no funnel asymmetry ($z=-0.624$; $p=0.533$), suggesting the absence of small-study or publication bias. The \textbf{trim-and-fill procedure} imputes zero missing studies; the bias-adjusted estimate remains essentially unchanged at $g = 0.273$, (\;95\% $[0.018, 0.528]$), corroborating the robustness of the performance benefit.
\textbf{Influence diagnostics:} One observation shows a high studentized residual ($r = 2.933$), but its removal reduces $\tau^{2}$ only from 0.3261 to 0.2271 and leaves $g$ within the original confidence limits. All other influence metrics (DFFITS, Cook's $D$, hat values, cov.r) remain below conventional thresholds. 

\subsection{RQ2b: Effect of Human-GenAI Collaboration on Creative Diversity}
\label{sec:CreativeDiversity}

In \textbf{RQ2b}, we aim to answer: \textit{How diverse are ideas generated by humans when supported by GenAI compared to humans without GenAI support}. Here, we analyze all studies comparing the creative performance of humans with vs. without GenAI support in regard to creative diversity. Note that creative diversity is measured only in a few studies (i.e., we have 6 observations with effect sizes), because of which the meta-analytic results below are subject to a small sample size. 

\begin{figure}[H]
  \centering
  \includegraphics[width=.65\linewidth]{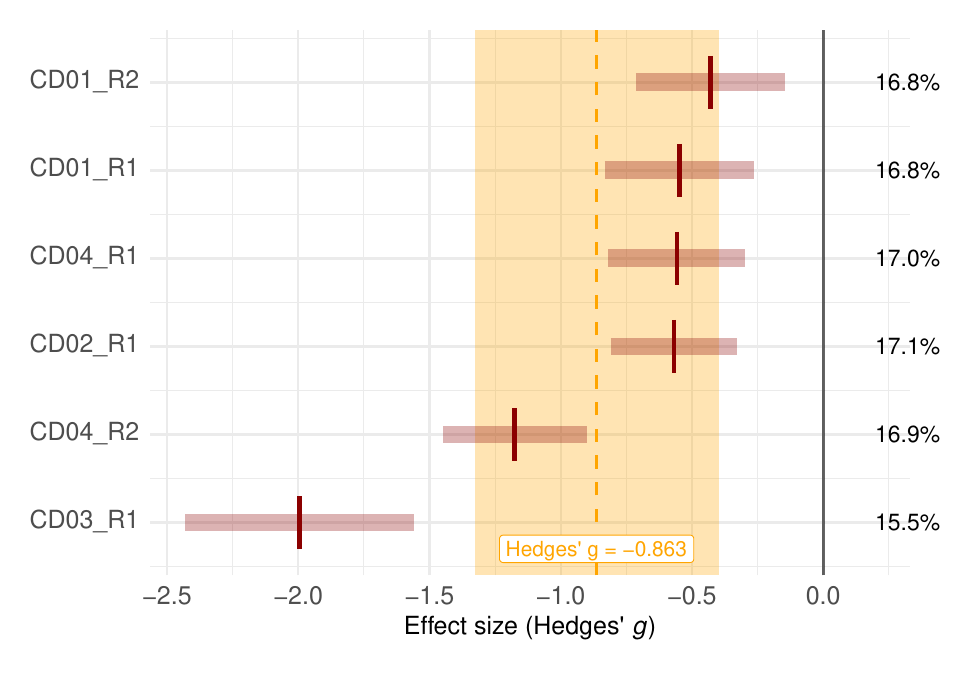}
  \caption{\textbf{Pooled effects (RQ2b).} Forest plot summarizing the Hedges' $g$ effect sizes and 95\% confidence intervals for creative diversity at the observation level (treatment: human-GenAI vs. control: Human alone), across six individual experiments from four studies. Out of the 127 observations, $n = 6$ observations (participants $m = 1017$) quantify differences in creative diversity between human and human-GenAI collaboration. Each line is one replication's estimate (the weight is shown at the right). The overall effect size of $g = -0.863$ indicates that GenAI assistance leads to ideas that are less diverse. The vertical line at $g = 0$ marks the null; points to the right favor the GenAI-assisted treatment. The bars are the estimated effect sizes, and the whiskers are the 95\% CIs. The orange dashed line is the mean pooled effect size and the orange shaded area is its 95\% CI.}
  \label{fig:RQ2b_pooled}
\end{figure}

\textbf{Pooled effect:} As shown in Figure~\ref{fig:RQ2b_pooled}, the effect of human–AI collaboration on diversity yields a pooled Hedges'\,$g = -0.863$ (95\% CI:\,[\,$-1.328, -0.398$], $p < 0.001$). Still, heterogeneity is severe ($Q_{5}=51.69$; $p<0.001$; $I^{2}=93.70\%$; $\tau^{2}=0.310$). Each study carries roughly 16\% to 17\% weight. Nevertheless, we find a consistent and statistically meaningful reduction in idea diversity when GenAI joins the team. 

\textbf{Robustness:} We again performed a leave-one-out sensitivity analysis. Sequential deletion keeps the pooled estimate between $g = -0.655$ and $g = -0.952$; every 95\% CI excludes zero, and $I^{2}$ never falls below 78.21\%. The diversity-reducing effect is therefore robust to the removal of any single study, albeit heterogeneity persists. 

\subsubsection{Heterogeneity Analysis}
\label{sec:CreativeDiversity_Moderator_GenAI}

Due to the small sample size and limited variation in experimental designs across studies, we restrict our focus to the following subgroup analysis around participants (i.e., we cannot perform subgroup analyses for GenAI models or task types as above).

\textbf{Participants.} Compared with laypeople samples, academic cohorts show a stronger reduction in diversity ($g = -1.260$; 95\% CI: $[-2.340, -0.187]$; $p = 0.021$), while business populations exhibit a non-significant shift ($g = -0.866$; 95\% CI: $[-1.930, 0.197]$; $p = 0.110$). Participant background thus moderates the magnitude of the negative diversity effect. 

\subsubsection{Robustness}

\textbf{Funnel plot \& Egger's test:} Egger's test and trim-and-fill were not executed due to the number of studies. \textbf{Influence diagnostics.} One outlier (study \#4) shows a studentized residual of $-3.662$ and Cook's $D = 0.766$, excising it lowers $\tau^{2}$ from 0.310 to 0.067 and $Q_{E}$ from 51.69 to 17.87, but the pooled effect remains negative and significant ($g = -0.656$; 95\% CI:\,$[\,-0.913, -0.398]$). Thus, although the outlier inflates heterogeneity, it does not alter the conclusion. 

\section{Discussion}
\label{sec:discussion}

\subsection{Summary of Key Findings}

The fragmented state of the literature in human-GenAI co-creativity research raises two central questions: \emph{How creative are the ideas generated by GenAI? And to what extent can GenAI support humans in producing ideas that are both creative and diverse?} For our meta-analysis, we screened 691 records and finally included 28 studies (with $m=8214$ participants) comparing creative performance along three dimensions: (i)~how creative GenAI is, (ii)~how creative human-GenAI collaboration is vs. human-only performance, and (iii)~the effect of human-GenAI collaboration on idea diversity.

\begin{table}[H]
\centering
\footnotesize
\begin{tabular}{@{}llcc@{}}
  \toprule
RQ  & Description                        & Hedges' $g$       & Supported? \\ 
  \midrule
RQ1 & Creative performance of GenAI (compared to human)            & $-0.05$ (ns)      & \textcolor{BrickRed}{\ding{55}} \\ 
RQ2a & Creative performance of human-GenAI (compared to human only)            & $0.27^{*}$      & \textcolor{ForestGreen}{\ding{51}} \\ 
RQ2b & Diversity of ideas in human-GenAI (compared to human only)              & $-0.86^{***}$      & \textcolor{ForestGreen}{\ding{51}} \\ 
  \bottomrule
\end{tabular}
\caption{\textbf{Key findings from our meta-analysis.} Reported are Hedges' $g$ for each anaysis. \textcolor{ForestGreen}{\ding{51}} indicates $p< 0.001$, \textcolor{BrickRed}{\ding{55}} indicates non-significant. Significance levels: $^{***}$: $p < 0.001$; $^{**}$: $p < 0.01$; $^{*}$: $p < 0.05$.}
\label{tab:rq_overview}
\end{table}

\textbf{Finding for RQ1: Parity rather than super-human creativity.} GenAI matches the average human creative output ($g \approx -0.05$; see Table~\ref{tab:rq_overview}). A moderate advantage ($g \approx 0.499$) emerges only for specific GenAI models and is largely attributable to a subset of studies using GPT-4. Hence, any observed superiority is tied to specific models, rather than being a general property of LLMs. This result is also in line with studies explicitly comparing the creative performance of different models, such as GPT-3 and GPT-4 \cite{Koivisto.2023}. Nevertheless, with the growing number of research using recent LLMs with reasoning capabilities, we may observe a gradual shift in average performance benchmarks. 

\textbf{Finding to RQ2a. Modest but robust gains from collaboration.} Human-GenAI teams deliver a small yet robust boost in creative performance ($g\approx 0.27$) that holds across different models, tasks, and participant backgrounds. While our meta-analysis focused on comparing human-GenAI collaboration with human-only performance (rather than contrasting human-GenAI teams with GenAI-only outputs), recent studies examining the latter report no significant performance gains from adding humans to GenAI-generated content  \cite{Lee.2024}. Together, humans using GenAI therefore show higher creative performance than humans without GenAI support.

\textbf{Finding to RQ2b. Decrease in idea diversity.}  Collaboration with GenAI is associated with a significant decrease in idea diversity (pooled $g \approx -0.86$), indicating a potential homogenization effect of the AI. This finding strengthens the impression that collaboration with LLMs might improve individual performance while showing detrimental effects on group-level \cite{Doshi.2024b}. It remains unclear whether this effect generalizes beyond GPT‑4 and text-based tasks, as the analysis aimed at understanding \textbf{RQ2b} was based on limited data.
 
\textbf{Heterogeneity in our findings}. Our heterogeneity analysis reveals three key insights. First, GenAI performs better on simple, standardized creativity tests---such as the alternative uses task (AUT) and consequences task (CT)---than on more complex, elaborative tasks like creative writing. Second, creative performance varies by model: newer models outperform older ones when generating ideas without human involvement, but, in human–AI collaboration, GPT-3.5 consistently improves creative output, while GPT-4 shows no significant added benefit. Third, laypeople particularly benefit from collaborating with GenAI, yet they tend to be outperformed when directly competing against it.

\subsection{Implications}

\textbf{Practical implications:} Our research suggests substantial potential for organizations to integrate GenAI into creative workflows---but only by pairing it with employees to enhance ideation processes. These benefits also imply a shift in individual creative potential: as GenAI can augment human creativity, especially lower innate creative capabilities might be compensated \cite{Doshi.2024b}. Yet, the observed gains in creativity could come at a cost, namely, due to reduced idea diversity. As a result, users collaborating with GenAI tend to produce more homogeneous outputs in some settings. This trade-off is especially problematic in contexts where ideational breadth is essential, such as where companies search for ``out-of-the-box'' ideas or in open innovation and crowd ideation tasks \cite{Boudreau.2013}. Thus, while human-GenAI teaming may suffice for boosting individual creativity, practitioners will need to be careful when scaling GenAI use with the intention of promoting collective creativity.

Further, current GenAI models often tend to recombine familiar elements \cite{Lee.2024} across contexts rather than generating radically novel ideas, which may lead to more incremental innovation. This highlights the importance of human agency in co-creation processes to counteract potential convergence effects---where repeated exposure to similar outputs leads to reduced variability---and highlights the need for human-GenAI systems that actively support diversity.

\textbf{Theoretical implications:} Our study reveals that the creative impact of GenAI is highly contingent on contextual factors. A high between-study heterogeneity $I^{2}$ >80\% demonstrates that creativity augmentation cannot be treated as a general affordance of GenAI alone. However, the effectiveness mechanisms of GenAI in creativity remain unclear. Importantly, our heterogeneity analysis identifies only the choice of the GenAI model as a consistent moderator that explains higher creativity levels. 

Given that the benefit from human-GenAI collaboration in creative tasks is robust for both laypeople and domain experts, it is likely that the performance gain is due to a generic facilitation mechanism (e.g. faster drafting and broader ideation, reduced cognitive load) rather than task-specific benefits from augmentation. Previous evidence indicates that co-creation tasks are particularly beneficial for human-GenAI collaboration when there are inherent synergies between humans and GenAI systems, allowing each to leverage their respective strengths---unlike in many analytical, decision-making contexts  \cite{JonathanA.Plucker.2004}.

\subsection{Limitations}
\label{sec:limitations}

Our meta-analysis has limitations due to the current state of research and technological progress. First, some of the included studies are not yet peer-reviewed. Nevertheless, we still chose to include them to provide a timely synthesis of empirical findings, including recent ones. Second, the findings depend on the choice of GenAI models. Given the fast-evolving capabilities of GenAI, it is likely that newer models, including improved reasoning capabilities or different fine-tuning strategies, may yield different outcomes. Third, many studies use simple, well-known tasks for benchmarking, such as the so-called alternative uses test, which asks participants to generate creative uses for common objects. Such tasks may have been part of the training data in GenAI models, which is a known issue in GenAI research \cite{Vaccaro.2024} and underscores the need for devising novel tasks that preserve experimental rigor while minimizing overlap with model training data.

\subsection{Future Research}
\label{sec:research_agenda}

To guide future work, we identified salient gaps around the use of GenAI for creativity systems in HCI/CSCW research based on our structured literature review and propose three key research directions.

\textbf{Research direction 1: \emph{More relevant, real-world-inspired study designs:}} The majority of studies focus on simplified creativity tasks such as the alternative uses test (AUT) \cite{Xu.662024}. While these tasks draw on established scales from psychological research to assess creativity, they fall short of capturing the complexity of real-world settings that involve creative thinking. In contrast, there are only a few studies that test GenAI in real-world situations such as work settings (e.g., \cite{Chen.2024}), or with creative tasks beyond text, such as images, audio, or code.

Future studies could focus on more realistic scenarios for a higher ecological validity and further explore context-sensitive moderators for more targeted creativity interventions. For example, to enhance relevance to CSCW and HCI applications in business contexts, future studies could investigate settings where tacit knowledge plays a critical role, or where creative tasks must account for specific organizational contexts---such as operational constraints or strategic alignment. One promising direction would be to examine how GenAI can support ideation in tasks like developing new business models that align with a company's brand strategy or existing capabilities of that company. Moreover, all of the existing studies were performed at a single time point; thus, longitudinal designs are needed to examine how collaboration patterns and creativity effects evolve over time---especially as users gain fluency in leveraging GenAI within creative tasks.

\textbf{Research direction 2: \emph{Understand psychological mechanisms:}} Existing research has primarily focused on the outcomes of GenAI-assisted creativity, typically through rigorous A/B experiments. Yet, prior research has largely overlooked the underlying psychological mechanisms that drive the effects. For example, it is unclear whether gains in GenAI-assisted creativity come from broader ideation (e.g., generating more ideas due to increased speed or reduced distractions), deeper engagement (e.g., more intensive thinking due to reduced cognitive load), or heightened motivation (e.g., playful interaction with the system). 

To develop a more nuanced understanding of human-GenAI co-creation, future work should explore potential psychological mediators and moderators, such as cognitive effort, user agency, trust, task framing, and participants personality traits. Uncovering such cognitive processes is relevant to identifying how human and machine capabilities can be combined to foster synergistic collaboration. Ultimately, such insights could inform frameworks with a notion of `appropriate reliance' (see \cite{Schemmer.2023} for an overview on the notion of appropriate reliance in AI advice), that is, when it is most beneficial for an idea to originate from the human, the GenAI system, or through their collaboration.
    
\textbf{Research direction 3: \emph{Explore CSCW/HCI design choices to elicit creative thinking:}} The majority of studies identified in our literature review rely on relatively naive GenAI setups---often based on simple one-shot prompts (e.g., as in \cite{Koivisto.2023}) or by providing participants with unmodified, off-the-shelf LLMs (e.g., as in \cite{Lee.2024}). In contrast, studies that devise and compare different design choices are largely absent. Future studies could thus explore more elaborate workflows (e.g., critique-refine, suggestions-refine) that reflect an iterative creation process. Eventually, this could help to integrate GenAI tools into creativity systems through effective, user-friendly interfaces that allow for natural interactions, as well as to build GenAI agents that are effective at supporting creative thinking tasks.

\section{Conclusion}
\label{sec:conclusion}

Our meta-analysis shows that GenAI and humans show similar creative performance on average, with moderate advantages emerging mainly for GPT-4. In contrast, human-GenAI collaboration shows small but consistent gains in creative output across tasks and contexts. However, collaboration with GenAI reduces the diversity of ideas, indicating a risk of creative outputs that could become more homogeneous. Our meta-analysis also highlights important gaps in the literature, which provide interesting avenues for future research to understand psychological mechanisms of human-GenAI augmentation and identify drivers for how GenAI systems can successfully elicit creative thinking.

\newpage
\bibliographystyle{ACM-Reference-Format}
\bibliography{library}


\begin{thebibliography}{73}


\ifx \showCODEN    \undefined \def \showCODEN     #1{\unskip}     \fi
\ifx \showISBNx    \undefined \def \showISBNx     #1{\unskip}     \fi
\ifx \showISBNxiii \undefined \def \showISBNxiii  #1{\unskip}     \fi
\ifx \showISSN     \undefined \def \showISSN      #1{\unskip}     \fi
\ifx \showLCCN     \undefined \def \showLCCN      #1{\unskip}     \fi
\ifx \shownote     \undefined \def \shownote      #1{#1}          \fi
\ifx \showarticletitle \undefined \def \showarticletitle #1{#1}   \fi
\ifx \showURL      \undefined \def \showURL       {\relax}        \fi
\providecommand\bibfield[2]{#2}
\providecommand\bibinfo[2]{#2}
\providecommand\natexlab[1]{#1}
\providecommand\showeprint[2][]{arXiv:#2}

\bibitem[Aru(2025)]%
        {Aru.2025}
\bibfield{author}{\bibinfo{person}{Jaan Aru}.} \bibinfo{year}{2025}\natexlab{}.
\newblock \showarticletitle{Artificial intelligence and the internal processes of creativity}.
\newblock \bibinfo{journal}{\emph{The Journal of Creative Behavior}} (\bibinfo{year}{2025}).
\newblock
\showISSN{0022-0175}
\href{https://doi.org/10.1002/jocb.1530}{doi:\nolinkurl{10.1002/jocb.1530}}


\bibitem[Ashkinaze et~al\mbox{.}(2024)]%
        {Ashkinaze.1242024}
\bibfield{author}{\bibinfo{person}{Joshua Ashkinaze}, \bibinfo{person}{Julia Mendelsohn}, \bibinfo{person}{Li Qiwei}, \bibinfo{person}{Ceren Budak}, {and} \bibinfo{person}{Eric Gilbert}.} \bibinfo{year}{2024}\natexlab{}.
\newblock \bibinfo{title}{How AI ideas affect the creativity, diversity, and evolution of human ideas: Evidence from a large, dynamic experiment}.
\newblock
\urldef\tempurl%
\url{https://arxiv.org/abs/2401.13481}
\showURL{%
\tempurl}


\bibitem[Borenstein et~al\mbox{.}(2021)]%
        {Borenstein.2021}
\bibfield{author}{\bibinfo{person}{Michael Borenstein}, \bibinfo{person}{Larry~V. Hedges}, \bibinfo{person}{Julian P.~T. Higgins}, {and} \bibinfo{person}{Hannah~R. Rothstein}.} \bibinfo{year}{2021}\natexlab{}.
\newblock \bibinfo{booktitle}{\emph{Introduction to meta-analysis}}.
\newblock \bibinfo{publisher}{{John Wiley {\&} Sons, Ltd}}, \bibinfo{address}{West Sussex, UK}.
\newblock
\showISBNx{1119558387}
\href{https://doi.org/10.1002/9780470743386}{doi:\nolinkurl{10.1002/9780470743386}}


\bibitem[Boudreau and Lakhani(2013)]%
        {Boudreau.2013}
\bibfield{author}{\bibinfo{person}{Kevin~J. Boudreau} {and} \bibinfo{person}{Karim~R. Lakhani}.} \bibinfo{year}{2013}\natexlab{}.
\newblock \showarticletitle{Using the crowd as an innovation partner}.
\newblock \bibinfo{journal}{\emph{Harvard Business Review}} \bibinfo{volume}{91}, \bibinfo{number}{4} (\bibinfo{year}{2013}), \bibinfo{pages}{60--69}.
\newblock
\showISSN{0017-8012}


\bibitem[Boussioux et~al\mbox{.}(2024)]%
        {Boussioux.2024}
\bibfield{author}{\bibinfo{person}{L{\'e}onard Boussioux}, \bibinfo{person}{Jacqueline~N. Lane}, \bibinfo{person}{Miaomiao Zhang}, \bibinfo{person}{Vladimir Jacimovic}, {and} \bibinfo{person}{Karim~R. Lakhani}.} \bibinfo{year}{2024}\natexlab{}.
\newblock \showarticletitle{The crowdless future? Generative AI and creative problem-solving}.
\newblock \bibinfo{journal}{\emph{Organization Science}} \bibinfo{volume}{35}, \bibinfo{number}{5} (\bibinfo{year}{2024}), \bibinfo{pages}{1589--1607}.
\newblock
\href{https://doi.org/10.1287/orsc.2023.18430}{doi:\nolinkurl{10.1287/orsc.2023.18430}}


\bibitem[Brynjolfsson et~al\mbox{.}(2025)]%
        {Brynjolfsson.2025}
\bibfield{author}{\bibinfo{person}{Erik Brynjolfsson}, \bibinfo{person}{Danielle Li}, {and} \bibinfo{person}{Lindsey Raymond}.} \bibinfo{year}{2025}\natexlab{}.
\newblock \showarticletitle{Generative AI at work}.
\newblock \bibinfo{journal}{\emph{The Quarterly Journal of Economics}} \bibinfo{volume}{140}, \bibinfo{number}{2} (\bibinfo{year}{2025}), \bibinfo{pages}{889--942}.
\newblock
\showISSN{0033-5533}
\href{https://doi.org/10.1093/qje/qjae044}{doi:\nolinkurl{10.1093/qje/qjae044}}


\bibitem[Castelo et~al\mbox{.}(2024)]%
        {Castelo.2024}
\bibfield{author}{\bibinfo{person}{Noah Castelo}, \bibinfo{person}{Zsolt Katona}, \bibinfo{person}{Peiyao Li}, {and} \bibinfo{person}{Miklos Sarvary}.} \bibinfo{year}{2024}\natexlab{}.
\newblock \showarticletitle{How AI outperforms humans at creative idea generation}.
\newblock \bibinfo{journal}{\emph{Available at SSRN 4751779}} (\bibinfo{year}{2024}).
\newblock
\href{https://doi.org/10.2139/ssrn.4751779}{doi:\nolinkurl{10.2139/ssrn.4751779}}


\bibitem[Charness and Grieco(2024)]%
        {Charness.2024}
\bibfield{author}{\bibinfo{person}{Gary Charness} {and} \bibinfo{person}{Daniela Grieco}.} \bibinfo{year}{2024}\natexlab{}.
\newblock \showarticletitle{Creativity and AI}.
\newblock \bibinfo{journal}{\emph{Available at SSRN 4686415}} (\bibinfo{year}{2024}).
\newblock
\href{https://doi.org/10.2139/ssrn.4686415}{doi:\nolinkurl{10.2139/ssrn.4686415}}


\bibitem[Chen and Chan(2024)]%
        {Chen.2024}
\bibfield{author}{\bibinfo{person}{Zenan Chen} {and} \bibinfo{person}{Jason Chan}.} \bibinfo{year}{2024}\natexlab{}.
\newblock \showarticletitle{Large language model in creative work: The role of collaboration modality and user expertise}.
\newblock \bibinfo{journal}{\emph{Management Science}} \bibinfo{volume}{70}, \bibinfo{number}{12} (\bibinfo{year}{2024}), \bibinfo{pages}{9101--9117}.
\newblock
\showISSN{0025-1909}
\href{https://doi.org/10.1287/mnsc.2023.03014}{doi:\nolinkurl{10.1287/mnsc.2023.03014}}


\bibitem[Cohen(2013)]%
        {Cohen.2013}
\bibfield{author}{\bibinfo{person}{Jacob Cohen}.} \bibinfo{year}{2013}\natexlab{}.
\newblock \bibinfo{booktitle}{\emph{Statistical power analysis for the behavioral sciences} (\bibinfo{edition}{2nd} ed.)}.
\newblock \bibinfo{publisher}{Routledge}, \bibinfo{address}{New York}.
\newblock
\showISBNx{9780203771587}
\href{https://doi.org/10.4324/9780203771587}{doi:\nolinkurl{10.4324/9780203771587}}


\bibitem[Cui et~al\mbox{.}(2024)]%
        {Cui.2024}
\bibfield{author}{\bibinfo{person}{Zheyuan~Kevin Cui}, \bibinfo{person}{Mert Demirer}, \bibinfo{person}{Sonia Jaffe}, \bibinfo{person}{Leon Musolff}, \bibinfo{person}{Sida Peng}, {and} \bibinfo{person}{Tobias Salz}.} \bibinfo{year}{2024}\natexlab{}.
\newblock \showarticletitle{The effects of generative ai on high skilled work: Evidence from three field experiments with software developers}.
\newblock \bibinfo{journal}{\emph{Available at SSRN 4945566}} (\bibinfo{year}{2024}).
\newblock
\href{https://doi.org/10.2139/ssrn.4945566}{doi:\nolinkurl{10.2139/ssrn.4945566}}


\bibitem[{de Vicente-Yag{\"u}e-Jara} et~al\mbox{.}(2023)]%
        {deVicenteYagueJaraMarinezOlivia.2023}
\bibfield{author}{\bibinfo{person}{Mar{\'i}a-Isabel {de Vicente-Yag{\"u}e-Jara}}, \bibinfo{person}{Olivia L{\'o}pez-Mart{\'i}nez}, \bibinfo{person}{Ver{\'o}nica Navarro-Navarro}, {and} \bibinfo{person}{Francisco Cu{\'e}llar-Santiago}.} \bibinfo{year}{2023}\natexlab{}.
\newblock \showarticletitle{Writing, creativity, and artificial intelligence: ChatGPT in the university context}.
\newblock \bibinfo{journal}{\emph{Comunicar: Media Education Research Journal}} \bibinfo{volume}{31}, \bibinfo{number}{77} (\bibinfo{year}{2023}), \bibinfo{pages}{45--54}.
\newblock
\href{https://doi.org/10.3916/C77-2023-04}{doi:\nolinkurl{10.3916/C77-2023-04}}


\bibitem[Deeks et~al\mbox{.}(2019)]%
        {Deeks.2019}
\bibfield{author}{\bibinfo{person}{Jonathan~J. Deeks}, \bibinfo{person}{Julian P.~T. Higgins}, \bibinfo{person}{Douglas~G. Altman}, {and} \bibinfo{person}{{on behalf of the Cochrane Statistical Methods Group}}.} \bibinfo{year}{2019}\natexlab{}.
\newblock \showarticletitle{Analysing data and undertaking meta-analyses: 10}.
\newblock In \bibinfo{booktitle}{\emph{Cochrane Handbook for Systematic Reviews of Interventions}}. \bibinfo{publisher}{{John Wiley {\&} Sons, Ltd}}, \bibinfo{pages}{241--284}.
\newblock
\showISBNx{9781119536604}
\href{https://doi.org/10.1002/9781119536604.ch10}{doi:\nolinkurl{10.1002/9781119536604.ch10}}


\bibitem[DerSimonian and Laird(1986)]%
        {DerSimonian.1986}
\bibfield{author}{\bibinfo{person}{Rebecca DerSimonian} {and} \bibinfo{person}{Nan Laird}.} \bibinfo{year}{1986}\natexlab{}.
\newblock \showarticletitle{Meta-analysis in clinical trials}.
\newblock \bibinfo{journal}{\emph{Controlled Clinical Trials}} \bibinfo{volume}{7}, \bibinfo{number}{3} (\bibinfo{year}{1986}), \bibinfo{pages}{177--188}.
\newblock
\showISSN{0197-2456}
\href{https://doi.org/10.1016/0197-2456(86)90046-2}{doi:\nolinkurl{10.1016/0197-2456(86)90046-2}}


\bibitem[Doshi et~al\mbox{.}(2024)]%
        {Doshi.2024}
\bibfield{author}{\bibinfo{person}{Anil~R. Doshi}, \bibinfo{person}{Sen Chai}, {and} \bibinfo{person}{Matthias Troebinger}.} \bibinfo{year}{2024}\natexlab{}.
\newblock \showarticletitle{How experience moderates the impact of generative AI ideas on the research process}.
\newblock \bibinfo{journal}{\emph{Available at SSRN 5013086}} (\bibinfo{year}{2024}).
\newblock
\href{https://doi.org/10.2139/ssrn.5013086}{doi:\nolinkurl{10.2139/ssrn.5013086}}


\bibitem[Doshi and {Oliver P. Hauser}(2024)]%
        {Doshi.2024b}
\bibfield{author}{\bibinfo{person}{Anil~R. Doshi} {and} \bibinfo{person}{{Oliver P. Hauser}}.} \bibinfo{year}{2024}\natexlab{}.
\newblock \showarticletitle{Generative AI enhances individual creativity but reduces the collective diversity of novel content}.
\newblock \bibinfo{journal}{\emph{Science Advances}} \bibinfo{volume}{10}, \bibinfo{number}{28} (\bibinfo{year}{2024}), \bibinfo{pages}{eadn5290}.
\newblock
\href{https://doi.org/10.1126/sciadv.adn5290}{doi:\nolinkurl{10.1126/sciadv.adn5290}}


\bibitem[Dunlap et~al\mbox{.}(1996)]%
        {Dunlap.1996}
\bibfield{author}{\bibinfo{person}{William~P. Dunlap}, \bibinfo{person}{Jose~M. Cortina}, \bibinfo{person}{Joel~B. Vaslow}, {and} \bibinfo{person}{Michael~J. Burke}.} \bibinfo{year}{1996}\natexlab{}.
\newblock \showarticletitle{Meta-analysis of experiments with matched groups or repeated measures designs}.
\newblock \bibinfo{journal}{\emph{Psychological Methods}} \bibinfo{volume}{1}, \bibinfo{number}{2} (\bibinfo{year}{1996}), \bibinfo{pages}{170--177}.
\newblock
\showISSN{1082-989X}
\href{https://doi.org/10.1037/1082-989X.1.2.170}{doi:\nolinkurl{10.1037/1082-989X.1.2.170}}


\bibitem[Duval and Tweedie(2000)]%
        {Duval.2000}
\bibfield{author}{\bibinfo{person}{Sue Duval} {and} \bibinfo{person}{Richard Tweedie}.} \bibinfo{year}{2000}\natexlab{}.
\newblock \showarticletitle{Trim and fill: a simple funnel-plot--based method of testing and adjusting for publication bias in meta-analysis}.
\newblock \bibinfo{journal}{\emph{Biometrics}} \bibinfo{volume}{56}, \bibinfo{number}{2} (\bibinfo{year}{2000}), \bibinfo{pages}{455--463}.
\newblock
\showISSN{0006-341X}
\href{https://doi.org/10.1111/j.0006-341x.2000.00455.x}{doi:\nolinkurl{10.1111/j.0006-341x.2000.00455.x}}


\bibitem[Egger et~al\mbox{.}(1997)]%
        {Egger.1997}
\bibfield{author}{\bibinfo{person}{Matthias Egger}, \bibinfo{person}{George~Davey Smith}, \bibinfo{person}{Martin Schneider}, {and} \bibinfo{person}{Christoph Minder}.} \bibinfo{year}{1997}\natexlab{}.
\newblock \showarticletitle{Bias in meta-analysis detected by a simple, graphical test}.
\newblock \bibinfo{journal}{\emph{bmj}} \bibinfo{volume}{315}, \bibinfo{number}{7109} (\bibinfo{year}{1997}), \bibinfo{pages}{629--634}.
\newblock
\href{https://doi.org/10.1136/bmj.315.7109.629}{doi:\nolinkurl{10.1136/bmj.315.7109.629}}


\bibitem[Eisenreich et~al\mbox{.}(2024)]%
        {AnjaEisenreich.2024}
\bibfield{author}{\bibinfo{person}{Anja Eisenreich}, \bibinfo{person}{Julian Just}, \bibinfo{person}{Daniela Gimenez-Jimenez}, {and} \bibinfo{person}{Johann F{\"u}ller}.} \bibinfo{year}{2024}\natexlab{}.
\newblock \showarticletitle{Revolution or inflated expectations? Exploring the impact of generative AI on ideation in a practical sustainability context}.
\newblock \bibinfo{journal}{\emph{Technovation}}  \bibinfo{volume}{138} (\bibinfo{year}{2024}), \bibinfo{pages}{103123}.
\newblock
\showISSN{01664972}
\href{https://doi.org/10.1016/j.technovation.2024.103123}{doi:\nolinkurl{10.1016/j.technovation.2024.103123}}


\bibitem[Feuerriegel et~al\mbox{.}(2024)]%
        {Feuerriegel.2024}
\bibfield{author}{\bibinfo{person}{Stefan Feuerriegel}, \bibinfo{person}{Jochen Hartmann}, \bibinfo{person}{Christian Janiesch}, {and} \bibinfo{person}{Patrick Zschech}.} \bibinfo{year}{2024}\natexlab{}.
\newblock \showarticletitle{Generative AI}.
\newblock \bibinfo{journal}{\emph{Business {\&} Information Systems Engineering}} \bibinfo{volume}{66}, \bibinfo{number}{1} (\bibinfo{year}{2024}), \bibinfo{pages}{111--126}.
\newblock
\showISSN{1867-0202}
\href{https://doi.org/10.1007/s12599-023-00834-7}{doi:\nolinkurl{10.1007/s12599-023-00834-7}}


\bibitem[Feuerriegel et~al\mbox{.}(2025)]%
        {Feuerriegel.2025}
\bibfield{author}{\bibinfo{person}{Stefan Feuerriegel}, \bibinfo{person}{Abdurahman Maarouf}, \bibinfo{person}{Dominik B{\"a}r}, \bibinfo{person}{Dominique Geissler}, \bibinfo{person}{Jonas Schweisthal}, \bibinfo{person}{Nicolas Pr{\"o}llochs}, \bibinfo{person}{Claire~E. Robertson}, \bibinfo{person}{Steve Rathje}, \bibinfo{person}{Jochen Hartmann}, \bibinfo{person}{Saif~M. Mohammad}, \bibinfo{person}{Oded Netzer}, \bibinfo{person}{Alexandra~A. Siegel}, \bibinfo{person}{Barbara Plank}, {and} \bibinfo{person}{Jay~J. {van Bavel}}.} \bibinfo{year}{2025}\natexlab{}.
\newblock \showarticletitle{Using natural language processing to analyse text data in behavioural science}.
\newblock \bibinfo{journal}{\emph{Nature Reviews Psychology}} \bibinfo{volume}{4}, \bibinfo{number}{2} (\bibinfo{year}{2025}), \bibinfo{pages}{96--111}.
\newblock
\showISSN{2731-0574}
\href{https://doi.org/10.1038/s44159-024-00392-z}{doi:\nolinkurl{10.1038/s44159-024-00392-z}}


\bibitem[G{\'o}mez-Rodr{\'i}guez and Williams(2023)]%
        {GomezRodriguez.10122023}
\bibfield{author}{\bibinfo{person}{Carlos G{\'o}mez-Rodr{\'i}guez} {and} \bibinfo{person}{Paul Williams}.} \bibinfo{year}{2023}\natexlab{}.
\newblock \bibinfo{title}{A confederacy of models: A comprehensive evaluation of LLMs on creative writing}.
\newblock
\urldef\tempurl%
\url{https://arxiv.org/abs/2310.08433v1}
\showURL{%
\tempurl}


\bibitem[Grassini and Koivisto(2025)]%
        {Grassini.2025}
\bibfield{author}{\bibinfo{person}{Simone Grassini} {and} \bibinfo{person}{Mika Koivisto}.} \bibinfo{year}{2025}\natexlab{}.
\newblock \showarticletitle{Artificial creativity? Evaluating AI against human performance in creative interpretation of visual stimuli}.
\newblock \bibinfo{journal}{\emph{International Journal of Human--Computer Interaction}} \bibinfo{volume}{41}, \bibinfo{number}{7} (\bibinfo{year}{2025}), \bibinfo{pages}{4037--4048}.
\newblock
\showISSN{1044-7318}
\href{https://doi.org/10.1080/10447318.2024.2345430}{doi:\nolinkurl{10.1080/10447318.2024.2345430}}


\bibitem[Gray et~al\mbox{.}(2019)]%
        {Gray.2019}
\bibfield{author}{\bibinfo{person}{Kurt Gray}, \bibinfo{person}{Stephen Anderson}, \bibinfo{person}{Eric~Evan Chen}, \bibinfo{person}{John~Michael Kelly}, \bibinfo{person}{Michael~S. Christian}, \bibinfo{person}{John Patrick}, \bibinfo{person}{Laura Huang}, \bibinfo{person}{Yoed~N. Kenett}, {and} \bibinfo{person}{Kevin Lewis}.} \bibinfo{year}{2019}\natexlab{}.
\newblock \showarticletitle{``Forward flow'': A new measure to quantify free thought and predict creativity}.
\newblock \bibinfo{journal}{\emph{American Psychologist}} \bibinfo{volume}{74}, \bibinfo{number}{5} (\bibinfo{year}{2019}), \bibinfo{pages}{539--554}.
\newblock
\href{https://doi.org/10.1037/amp0000391}{doi:\nolinkurl{10.1037/amp0000391}}


\bibitem[Grimes et~al\mbox{.}(2023)]%
        {Grimes.2023}
\bibfield{author}{\bibinfo{person}{Matthew Grimes}, \bibinfo{person}{Georg von Krogh}, \bibinfo{person}{Stefan Feuerriegel}, \bibinfo{person}{Floor Rink}, {and} \bibinfo{person}{Marc Gruber}.} \bibinfo{year}{2023}\natexlab{}.
\newblock \showarticletitle{From Scarcity to Abundance: Scholars and Scholarship~in an Age of Generative Artificial~Intelligence}.
\newblock \bibinfo{journal}{\emph{Academy of Management Journal}} \bibinfo{volume}{66}, \bibinfo{number}{6} (\bibinfo{year}{2023}), \bibinfo{pages}{1617--1624}.
\newblock
\href{https://doi.org/10.5465/amj.2023.4006}{doi:\nolinkurl{10.5465/amj.2023.4006}}


\bibitem[Guilford et~al\mbox{.}(1978)]%
        {Guilford.1978}
\bibfield{author}{\bibinfo{person}{Joy~Paul Guilford}, \bibinfo{person}{Paul~R. Christensen}, \bibinfo{person}{Philip~R. Merrifield}, {and} \bibinfo{person}{Robert~C. Wilson}.} \bibinfo{year}{1978}\natexlab{}.
\newblock \showarticletitle{Alternate uses}.
\newblock  (\bibinfo{year}{1978}).
\newblock
\href{https://doi.org/10.1037/t06443-000}{doi:\nolinkurl{10.1037/t06443-000}}


\bibitem[Herbold et~al\mbox{.}(2023)]%
        {Herbold.2023}
\bibfield{author}{\bibinfo{person}{Steffen Herbold}, \bibinfo{person}{Annette Hautli-Janisz}, \bibinfo{person}{Ute Heuer}, \bibinfo{person}{Zlata Kikteva}, {and} \bibinfo{person}{Alexander Trautsch}.} \bibinfo{year}{2023}\natexlab{}.
\newblock \showarticletitle{A large-scale comparison of human-written versus ChatGPT-generated essays}.
\newblock \bibinfo{journal}{\emph{Scientific Reports}} \bibinfo{volume}{13}, \bibinfo{number}{1} (\bibinfo{year}{2023}), \bibinfo{pages}{18617}.
\newblock
\href{https://doi.org/10.1038/s41598-023-45644-9}{doi:\nolinkurl{10.1038/s41598-023-45644-9}}


\bibitem[Hubert et~al\mbox{.}(2024)]%
        {Hubert.2024}
\bibfield{author}{\bibinfo{person}{Kent~F. Hubert}, \bibinfo{person}{Kim~N. Awa}, {and} \bibinfo{person}{Darya~L. Zabelina}.} \bibinfo{year}{2024}\natexlab{}.
\newblock \showarticletitle{The current state of artificial intelligence generative language models is more creative than humans on divergent thinking tasks}.
\newblock \bibinfo{journal}{\emph{Scientific Reports}} \bibinfo{volume}{14}, \bibinfo{number}{1} (\bibinfo{year}{2024}), \bibinfo{pages}{3440}.
\newblock
\href{https://doi.org/10.1038/s41598-024-53303-w}{doi:\nolinkurl{10.1038/s41598-024-53303-w}}


\bibitem[Ismayilzada et~al\mbox{.}(2024a)]%
        {Ismayilzada.10222024}
\bibfield{author}{\bibinfo{person}{Mete Ismayilzada}, \bibinfo{person}{Debjit Paul}, \bibinfo{person}{Antoine Bosselut}, {and} \bibinfo{person}{Lonneke {van der Plas}}.} \bibinfo{year}{2024}\natexlab{a}.
\newblock \bibinfo{title}{Creativity in AI: Progresses and challenges}.
\newblock
\urldef\tempurl%
\url{https://arxiv.org/abs/2410.17218v4}
\showURL{%
\tempurl}


\bibitem[Ismayilzada et~al\mbox{.}(2024b)]%
        {Ismayilzada.1142024}
\bibfield{author}{\bibinfo{person}{Mete Ismayilzada}, \bibinfo{person}{Claire Stevenson}, {and} \bibinfo{person}{Lonneke {van der Plas}}.} \bibinfo{year}{2024}\natexlab{b}.
\newblock \bibinfo{title}{Evaluating creative short story generation in humans and large language models}.
\newblock
\urldef\tempurl%
\url{https://arxiv.org/abs/2411.02316v5}
\showURL{%
\tempurl}


\bibitem[Jackson and Bowden(2016)]%
        {Jackson.2016}
\bibfield{author}{\bibinfo{person}{Dan Jackson} {and} \bibinfo{person}{Jack Bowden}.} \bibinfo{year}{2016}\natexlab{}.
\newblock \showarticletitle{Confidence intervals for the between-study variance in random-effects meta-analysis using generalised heterogeneity statistics: should we use unequal tails?}
\newblock \bibinfo{journal}{\emph{BMC Medical Research Methodology}} \bibinfo{volume}{16}, \bibinfo{number}{1} (\bibinfo{year}{2016}), \bibinfo{pages}{118}.
\newblock
\href{https://doi.org/10.1186/s12874-016-0219-y}{doi:\nolinkurl{10.1186/s12874-016-0219-y}}


\bibitem[{Jay A. Olson} et~al\mbox{.}(2021)]%
        {JayA.Olson.2021}
\bibfield{author}{\bibinfo{person}{{Jay A. Olson}}, \bibinfo{person}{{Johnny Nahas}}, \bibinfo{person}{{Denis Chmoulevitch}}, \bibinfo{person}{{Simon J. Cropper}}, {and} \bibinfo{person}{{Margaret E. Webb}}.} \bibinfo{year}{2021}\natexlab{}.
\newblock \showarticletitle{Naming unrelated words predicts creativity}.
\newblock \bibinfo{journal}{\emph{Proceedings of the National Academy of Sciences}} \bibinfo{volume}{118}, \bibinfo{number}{25} (\bibinfo{year}{2021}), \bibinfo{pages}{e2022340118}.
\newblock
\href{https://doi.org/10.1073/pnas.2022340118}{doi:\nolinkurl{10.1073/pnas.2022340118}}


\bibitem[{Jennifer Haase} and {Paul H.P. Hanel}(2023)]%
        {JenniferHaase.2023}
\bibfield{author}{\bibinfo{person}{{Jennifer Haase}} {and} \bibinfo{person}{{Paul H.P. Hanel}}.} \bibinfo{year}{2023}\natexlab{}.
\newblock \showarticletitle{Artificial muses: Generative artificial intelligence chatbots have risen to human-level creativity}.
\newblock \bibinfo{journal}{\emph{Journal of Creativity}} \bibinfo{volume}{33}, \bibinfo{number}{3} (\bibinfo{year}{2023}), \bibinfo{pages}{100066}.
\newblock
\showISSN{27133745}
\href{https://doi.org/10.1016/j.yjoc.2023.100066}{doi:\nolinkurl{10.1016/j.yjoc.2023.100066}}


\bibitem[{Jonathan A. Plucker} et~al\mbox{.}(2004)]%
        {JonathanA.Plucker.2004}
\bibfield{author}{\bibinfo{person}{{Jonathan A. Plucker}}, \bibinfo{person}{{Ronald A. Beghetto}}, {and} \bibinfo{person}{{Gayle T. Dow and}}.} \bibinfo{year}{2004}\natexlab{}.
\newblock \showarticletitle{Why isn't creativity more important to educational psychologists? Potentials, pitfalls, and future directions in creativity research}.
\newblock \bibinfo{journal}{\emph{Educational Psychologist}} \bibinfo{volume}{39}, \bibinfo{number}{2} (\bibinfo{year}{2004}), \bibinfo{pages}{83--96}.
\newblock
\showISSN{0046-1520}
\href{https://doi.org/10.1207/s15326985ep3902{\textunderscore }1}{doi:\nolinkurl{10.1207/s15326985ep3902{\textunderscore }1}}


\bibitem[Kapoor and Kumar(2025)]%
        {Kapoor.0}
\bibfield{author}{\bibinfo{person}{Anuj Kapoor} {and} \bibinfo{person}{Madhav Kumar}.} \bibinfo{year}{2025}\natexlab{}.
\newblock \showarticletitle{Frontiers: Generative AI and personalized video advertisements}.
\newblock \bibinfo{journal}{\emph{Marketing Science}} (\bibinfo{year}{2025}).
\newblock
\showISSN{0732-2399}
\href{https://doi.org/10.1287/mksc.2023.0494}{doi:\nolinkurl{10.1287/mksc.2023.0494}}


\bibitem[Koivisto and Grassini(2023)]%
        {Koivisto.2023}
\bibfield{author}{\bibinfo{person}{Mika Koivisto} {and} \bibinfo{person}{Simone Grassini}.} \bibinfo{year}{2023}\natexlab{}.
\newblock \showarticletitle{Best humans still outperform artificial intelligence in a creative divergent thinking task}.
\newblock \bibinfo{journal}{\emph{Scientific Reports}} \bibinfo{volume}{13}, \bibinfo{number}{1} (\bibinfo{year}{2023}), \bibinfo{pages}{13601}.
\newblock
\href{https://doi.org/10.1038/s41598-023-40858-3}{doi:\nolinkurl{10.1038/s41598-023-40858-3}}


\bibitem[{Larry V. Hedges}(1981)]%
        {LarryV.Hedges.1981}
\bibfield{author}{\bibinfo{person}{{Larry V. Hedges}}.} \bibinfo{year}{1981}\natexlab{}.
\newblock \showarticletitle{Distribution theory for glass's estimator of effect size and related estimators}.
\newblock \bibinfo{journal}{\emph{Journal of Educational Statistics}} \bibinfo{volume}{6}, \bibinfo{number}{2} (\bibinfo{year}{1981}), \bibinfo{pages}{107--128}.
\newblock
\showISSN{0362-9791}
\href{https://doi.org/10.3102/10769986006002107}{doi:\nolinkurl{10.3102/10769986006002107}}


\bibitem[Lee and Chung(2024)]%
        {Lee.2024}
\bibfield{author}{\bibinfo{person}{Byung~Cheol Lee} {and} \bibinfo{person}{Jaeyeon Chung}.} \bibinfo{year}{2024}\natexlab{}.
\newblock \showarticletitle{An empirical investigation of the impact of ChatGPT on creativity}.
\newblock \bibinfo{journal}{\emph{Nature Human Behaviour}} \bibinfo{volume}{8}, \bibinfo{number}{10} (\bibinfo{year}{2024}), \bibinfo{pages}{1906--1914}.
\newblock
\href{https://doi.org/10.1038/s41562-024-01953-1}{doi:\nolinkurl{10.1038/s41562-024-01953-1}}


\bibitem[{Mark A. Runco} and {Garrett J. Jaeger and}(2012)]%
        {MarkA.Runco.2012}
\bibfield{author}{\bibinfo{person}{{Mark A. Runco}} {and} \bibinfo{person}{{Garrett J. Jaeger and}}.} \bibinfo{year}{2012}\natexlab{}.
\newblock \showarticletitle{The standard definition of creativity}.
\newblock \bibinfo{journal}{\emph{Creativity Research Journal}} \bibinfo{volume}{24}, \bibinfo{number}{1} (\bibinfo{year}{2012}), \bibinfo{pages}{92--96}.
\newblock
\showISSN{1040-0419}
\href{https://doi.org/10.1080/10400419.2012.650092}{doi:\nolinkurl{10.1080/10400419.2012.650092}}


\bibitem[McGuire et~al\mbox{.}(2024)]%
        {McGuire.2024}
\bibfield{author}{\bibinfo{person}{Jack McGuire}, \bibinfo{person}{David de Cremer}, {and} \bibinfo{person}{Tim {van de Cruys}}.} \bibinfo{year}{2024}\natexlab{}.
\newblock \showarticletitle{Establishing the importance of co-creation and self-efficacy in creative collaboration with artificial intelligence}.
\newblock \bibinfo{journal}{\emph{Scientific Reports}} \bibinfo{volume}{14}, \bibinfo{number}{1} (\bibinfo{year}{2024}), \bibinfo{pages}{18525}.
\newblock
\href{https://doi.org/10.1038/s41598-024-69423-2}{doi:\nolinkurl{10.1038/s41598-024-69423-2}}


\bibitem[Mei et~al\mbox{.}(2025)]%
        {Mei.2025}
\bibfield{author}{\bibinfo{person}{Peidong Mei}, \bibinfo{person}{Deborah~N. Brewis}, \bibinfo{person}{Fortune Nwaiwu}, \bibinfo{person}{Deshan Sumanathilaka}, \bibinfo{person}{Fernando Alva-Manchego}, {and} \bibinfo{person}{Joanna Demaree-Cotton}.} \bibinfo{year}{2025}\natexlab{}.
\newblock \showarticletitle{If ChatGPT can do it, where is my creativity? Generative AI boosts performance but diminishes experience in creative writing}.
\newblock \bibinfo{journal}{\emph{Computers in Human Behavior: Artificial Humans}}  \bibinfo{volume}{4} (\bibinfo{year}{2025}), \bibinfo{pages}{100140}.
\newblock
\showISSN{29498821}
\href{https://doi.org/10.1016/j.chbah.2025.100140}{doi:\nolinkurl{10.1016/j.chbah.2025.100140}}


\bibitem[{Noah Bohren} et~al\mbox{.}(2024)]%
        {NoahBohren.2024}
\bibfield{author}{\bibinfo{person}{{Noah Bohren}}, \bibinfo{person}{{Rustamdjan Hakimov}}, {and} \bibinfo{person}{{Rafael Lalive}}.} \bibinfo{year}{2024}\natexlab{}.
\newblock \bibinfo{title}{Creative and strategic capabilities of Generative AI: Evidence from large-scale experiments: IZA Discussion Papers}.
\newblock
\urldef\tempurl%
\url{https://hdl.handle.net/10419/305744}
\showURL{%
\tempurl}


\bibitem[Orwig et~al\mbox{.}(2024)]%
        {Orwig.2024}
\bibfield{author}{\bibinfo{person}{William Orwig}, \bibinfo{person}{Emma~R. Edenbaum}, \bibinfo{person}{Joshua~D. Greene}, {and} \bibinfo{person}{Daniel~L. Schacter}.} \bibinfo{year}{2024}\natexlab{}.
\newblock \showarticletitle{The language of creativity: Evidence from humans and large language models}.
\newblock \bibinfo{journal}{\emph{The Journal of Creative Behavior}} \bibinfo{volume}{58}, \bibinfo{number}{1} (\bibinfo{year}{2024}), \bibinfo{pages}{128--136}.
\newblock
\showISSN{0022-0175}
\href{https://doi.org/10.1002/jocb.636}{doi:\nolinkurl{10.1002/jocb.636}}


\bibitem[Page et~al\mbox{.}(2021a)]%
        {Page.2021}
\bibfield{author}{\bibinfo{person}{Matthew~J. Page}, \bibinfo{person}{Joanne~E. McKenzie}, \bibinfo{person}{Patrick~M. Bossuyt}, \bibinfo{person}{Isabelle Boutron}, \bibinfo{person}{Tammy~C. Hoffmann}, \bibinfo{person}{Cynthia~D. Mulrow}, \bibinfo{person}{Larissa Shamseer}, \bibinfo{person}{Jennifer~M. Tetzlaff}, \bibinfo{person}{Elie~A. Akl}, \bibinfo{person}{Sally~E. Brennan}, \bibinfo{person}{Roger Chou}, \bibinfo{person}{Julie Glanville}, \bibinfo{person}{Jeremy~M. Grimshaw}, \bibinfo{person}{Asbj{\o}rn Hr{\'o}bjartsson}, \bibinfo{person}{Manoj~M. Lalu}, \bibinfo{person}{Tianjing Li}, \bibinfo{person}{Elizabeth~W. Loder}, \bibinfo{person}{Evan Mayo-Wilson}, \bibinfo{person}{Steve McDonald}, \bibinfo{person}{Luke~A. McGuinness}, \bibinfo{person}{Lesley~A. Stewart}, \bibinfo{person}{James Thomas}, \bibinfo{person}{Andrea~C. Tricco}, \bibinfo{person}{Vivian~A. Welch}, \bibinfo{person}{Penny Whiting}, {and} \bibinfo{person}{David Moher}.} \bibinfo{year}{2021}\natexlab{a}.
\newblock \bibinfo{title}{PRISMA 2020 Checklist}.
\newblock
\href{https://doi.org/10.1136/bmj.n71}{doi:\nolinkurl{10.1136/bmj.n71}}


\bibitem[Page et~al\mbox{.}(2021b)]%
        {Page.2021b}
\bibfield{author}{\bibinfo{person}{Matthew~J. Page}, \bibinfo{person}{Joanne~E. McKenzie}, \bibinfo{person}{Patrick~M. Bossuyt}, \bibinfo{person}{Isabelle Boutron}, \bibinfo{person}{Tammy~C. Hoffmann}, \bibinfo{person}{Cynthia~D. Mulrow}, \bibinfo{person}{Larissa Shamseer}, \bibinfo{person}{Jennifer~M. Tetzlaff}, \bibinfo{person}{Elie~A. Akl}, \bibinfo{person}{Sally~E. Brennan}, \bibinfo{person}{Roger Chou}, \bibinfo{person}{Julie Glanville}, \bibinfo{person}{Jeremy~M. Grimshaw}, \bibinfo{person}{Asbj{\o}rn Hr{\'o}bjartsson}, \bibinfo{person}{Manoj~M. Lalu}, \bibinfo{person}{Tianjing Li}, \bibinfo{person}{Elizabeth~W. Loder}, \bibinfo{person}{Evan Mayo-Wilson}, \bibinfo{person}{Steve McDonald}, \bibinfo{person}{Luke~A. McGuinness}, \bibinfo{person}{Lesley~A. Stewart}, \bibinfo{person}{James Thomas}, \bibinfo{person}{Andrea~C. Tricco}, \bibinfo{person}{Vivian~A. Welch}, \bibinfo{person}{Penny Whiting}, {and} \bibinfo{person}{David Moher}.} \bibinfo{year}{2021}\natexlab{b}.
\newblock \bibinfo{title}{PRISMA 2020 Flow Diagram}.
\newblock
\href{https://doi.org/10.1136/bmj.n71}{doi:\nolinkurl{10.1136/bmj.n71}}


\bibitem[Parjanen(2012)]%
        {Parjanen.2012}
\bibfield{author}{\bibinfo{person}{Satu Parjanen}.} \bibinfo{year}{2012}\natexlab{}.
\newblock \showarticletitle{Experiencing creativity in the organization: From individual creativity to collective creativity}.
\newblock \bibinfo{journal}{\emph{Interdisciplinary Journal of Information, Knowledge {\&} Management}}  \bibinfo{volume}{7} (\bibinfo{year}{2012}), \bibinfo{pages}{109--128}.
\newblock
\href{https://doi.org/10.28945/1580}{doi:\nolinkurl{10.28945/1580}}


\bibitem[Peterson and Brown(2005)]%
        {Peterson.2005}
\bibfield{author}{\bibinfo{person}{Robert~A. Peterson} {and} \bibinfo{person}{Steven~P. Brown}.} \bibinfo{year}{2005}\natexlab{}.
\newblock \showarticletitle{On the use of beta coefficients in meta-analysis}.
\newblock \bibinfo{journal}{\emph{Journal of Applied Psychology}} \bibinfo{volume}{90}, \bibinfo{number}{1} (\bibinfo{year}{2005}), \bibinfo{pages}{175--181}.
\newblock
\href{https://doi.org/10.1037/0021-9010.90.1.175}{doi:\nolinkurl{10.1037/0021-9010.90.1.175}}


\bibitem[Rosnow and Rosenthal(1996)]%
        {Rosnow.1996}
\bibfield{author}{\bibinfo{person}{Ralph~L. Rosnow} {and} \bibinfo{person}{Robert Rosenthal}.} \bibinfo{year}{1996}\natexlab{}.
\newblock \showarticletitle{Computing contrasts, effect sizes, and counternulls on other people's published data: General procedures for research consumers}.
\newblock \bibinfo{journal}{\emph{Psychological Methods}} \bibinfo{volume}{1}, \bibinfo{number}{4} (\bibinfo{year}{1996}), \bibinfo{pages}{331}.
\newblock
\showISSN{1082-989X}
\href{https://doi.org/10.1037/1082-989X.1.4.331}{doi:\nolinkurl{10.1037/1082-989X.1.4.331}}


\bibitem[Ruksakulpiwat et~al\mbox{.}(2024)]%
        {Ruksakulpiwat.2024}
\bibfield{author}{\bibinfo{person}{Suebsarn Ruksakulpiwat}, \bibinfo{person}{Lalipat Phianhasin}, \bibinfo{person}{Chitchanok Benjasirisan}, \bibinfo{person}{Kedong Ding}, \bibinfo{person}{Anuoluwapo Ajibade}, \bibinfo{person}{Ayanesh Kumar}, {and} \bibinfo{person}{Cassie Stewart}.} \bibinfo{year}{2024}\natexlab{}.
\newblock \showarticletitle{Assessing the efficacy of ChatGPT versus human researchers in identifying relevant studies on mHealth interventions for improving medication adherence in patients with ischemic stroke when conducting systematic reviews: Comparative analysis}.
\newblock \bibinfo{journal}{\emph{JMIR mHealth and uHealth}}  \bibinfo{volume}{12} (\bibinfo{year}{2024}), \bibinfo{pages}{e51526}.
\newblock
\href{https://doi.org/10.2196/51526}{doi:\nolinkurl{10.2196/51526}}


\bibitem[S{\ae}b{\o} and Brovold(2024)]%
        {Sb.332024}
\bibfield{author}{\bibinfo{person}{Solve S{\ae}b{\o}} {and} \bibinfo{person}{Helge Brovold}.} \bibinfo{year}{2024}\natexlab{}.
\newblock \bibinfo{title}{On the stochastics of human and artificial creativity}.
\newblock
\urldef\tempurl%
\url{https://arxiv.org/abs/2403.06996v1}
\showURL{%
\tempurl}


\bibitem[Schemmer et~al\mbox{.}(2022)]%
        {Schemmer.2022}
\bibfield{author}{\bibinfo{person}{Max Schemmer}, \bibinfo{person}{Patrick Hemmer}, \bibinfo{person}{Maximilian Nitsche}, \bibinfo{person}{Niklas K{\"u}hl}, {and} \bibinfo{person}{Michael V{\"o}ssing}.} \bibinfo{year}{2022}\natexlab{}.
\newblock \showarticletitle{A meta-analysis of the utility of explainable artificial intelligence in human-AI decision-making}. In \bibinfo{booktitle}{\emph{AAAI/ACM Conference on AI, Ethics, and Society}} \emph{(\bibinfo{series}{AIES})}.
\newblock
\showISBNx{9781450392471}
\href{https://doi.org/10.1145/3514094.3534128}{doi:\nolinkurl{10.1145/3514094.3534128}}


\bibitem[Schemmer et~al\mbox{.}(2023)]%
        {Schemmer.2023}
\bibfield{author}{\bibinfo{person}{Max Schemmer}, \bibinfo{person}{Niklas Kuehl}, \bibinfo{person}{Carina Benz}, \bibinfo{person}{Andrea Bartos}, {and} \bibinfo{person}{Gerhard Satzger}.} \bibinfo{year}{2023}\natexlab{}.
\newblock \showarticletitle{Appropriate reliance on AI advice: Conceptualization and the effect of explanations}. In \bibinfo{booktitle}{\emph{International Conference on Intelligent User Interfaces}} \emph{(\bibinfo{series}{IUI})}.
\newblock
\showISBNx{9798400701061}
\href{https://doi.org/10.1145/3581641.3584066}{doi:\nolinkurl{10.1145/3581641.3584066}}


\bibitem[Si et~al\mbox{.}(2024)]%
        {Si.962024}
\bibfield{author}{\bibinfo{person}{Chenglei Si}, \bibinfo{person}{Diyi Yang}, {and} \bibinfo{person}{Tatsunori Hashimoto}.} \bibinfo{year}{2024}\natexlab{}.
\newblock \bibinfo{title}{Can LLMs generate novel research ideas? A large-scale human study with 100+ NLP researchers}.
\newblock
\urldef\tempurl%
\url{https://arxiv.org/abs/2409.04109}
\showURL{%
\tempurl}


\bibitem[Song et~al\mbox{.}(2025)]%
        {Song.2025}
\bibfield{author}{\bibinfo{person}{Yu Song}, \bibinfo{person}{Longchao Huang}, \bibinfo{person}{Lanqin Zheng}, \bibinfo{person}{Mengya Fan}, {and} \bibinfo{person}{Zehao Liu}.} \bibinfo{year}{2025}\natexlab{}.
\newblock \showarticletitle{Interactions with generative AI chatbots: unveiling dialogic dynamics, students' perceptions, and practical competencies in creative problem-solving}.
\newblock \bibinfo{journal}{\emph{International Journal of Educational Technology in Higher Education}} \bibinfo{volume}{22}, \bibinfo{number}{1} (\bibinfo{year}{2025}), \bibinfo{pages}{12}.
\newblock
\showISSN{2365-9440}
\href{https://doi.org/10.1186/s41239-025-00508-2}{doi:\nolinkurl{10.1186/s41239-025-00508-2}}


\bibitem[SonicRim(2001)]%
        {SonicRim.2001}
\bibfield{author}{\bibinfo{person}{Liz~Sanders SonicRim}.} \bibinfo{year}{2001}\natexlab{}.
\newblock \showarticletitle{Collective creativity}.
\newblock \bibinfo{journal}{\emph{Design}} \bibinfo{volume}{6}, \bibinfo{number}{3} (\bibinfo{year}{2001}), \bibinfo{pages}{1--6}.
\newblock
\urldef\tempurl%
\url{http://www.echo.iat.sfu.ca/library/sanders_01_collective_creativity.pdf}
\showURL{%
\tempurl}


\bibitem[Sterne et~al\mbox{.}(2019)]%
        {Sterne.2019}
\bibfield{author}{\bibinfo{person}{Jonathan A.~C. Sterne}, \bibinfo{person}{Jelena Savovi{\'c}}, \bibinfo{person}{Matthew~J. Page}, \bibinfo{person}{Roy~G. Elbers}, \bibinfo{person}{Natalie~S. Blencowe}, \bibinfo{person}{Isabelle Boutron}, \bibinfo{person}{Christopher~J. Cates}, \bibinfo{person}{Hung-Yuan Cheng}, \bibinfo{person}{Mark~S. Corbett}, \bibinfo{person}{Sandra~M. Eldridge}, \bibinfo{person}{Jonathan~R. Emberson}, \bibinfo{person}{Miguel~A. Hern{\'a}n}, \bibinfo{person}{Sally Hopewell}, \bibinfo{person}{Asbj{\o}rn Hr{\'o}bjartsson}, \bibinfo{person}{Daniela~R. Junqueira}, \bibinfo{person}{Peter J{\"u}ni}, \bibinfo{person}{Jamie~J. Kirkham}, \bibinfo{person}{Toby Lasserson}, \bibinfo{person}{Tianjing Li}, \bibinfo{person}{Alexandra McAleenan}, \bibinfo{person}{Barnaby~C. Reeves}, \bibinfo{person}{Sasha Shepperd}, \bibinfo{person}{Ian Shrier}, \bibinfo{person}{Lesley~A. Stewart}, \bibinfo{person}{Kate Tilling}, \bibinfo{person}{Ian~R. White}, \bibinfo{person}{Penny~F. Whiting}, {and}
  \bibinfo{person}{Julian P.~T. Higgins}.} \bibinfo{year}{2019}\natexlab{}.
\newblock \showarticletitle{RoB 2: a revised tool for assessing risk of bias in randomised trials}.
\newblock \bibinfo{journal}{\emph{bmj}}  \bibinfo{volume}{366} (\bibinfo{year}{2019}), \bibinfo{pages}{4898}.
\newblock
\href{https://doi.org/10.1136/bmj.l4898}{doi:\nolinkurl{10.1136/bmj.l4898}}


\bibitem[Sun et~al\mbox{.}(2024)]%
        {Sun.1242024}
\bibfield{author}{\bibinfo{person}{Luning Sun}, \bibinfo{person}{Yuzhuo Yuan}, \bibinfo{person}{Yuan Yao}, \bibinfo{person}{Yanyan Li}, \bibinfo{person}{Hao Zhang}, \bibinfo{person}{Xing Xie}, \bibinfo{person}{Xiting Wang}, \bibinfo{person}{Fang Luo}, {and} \bibinfo{person}{David Stillwell}.} \bibinfo{year}{2024}\natexlab{}.
\newblock \bibinfo{title}{Large language models show both individual and collective creativity comparable to humans}.
\newblock
\urldef\tempurl%
\url{https://arxiv.org/abs/2412.03151}
\showURL{%
\tempurl}


\bibitem[Sun et~al\mbox{.}(2024)]%
        {Sun.2024}
\bibfield{author}{\bibinfo{person}{Yuan Sun}, \bibinfo{person}{Eunchae Jang}, \bibinfo{person}{Fenglong Ma}, {and} \bibinfo{person}{Ting Wang}.} \bibinfo{year}{2024}\natexlab{}.
\newblock \showarticletitle{Generative AI in the wild: Prospects, challenges, and strategies}. In \bibinfo{booktitle}{\emph{CHI Conference on Human Factors in Computing Systems}} \emph{(\bibinfo{series}{CHI})}.
\newblock
\showISBNx{9798400703300}
\href{https://doi.org/10.1145/3613904.3642160}{doi:\nolinkurl{10.1145/3613904.3642160}}


\bibitem[{Susannah B.F. Paletz} and {Kaiping Peng}(2008)]%
        {SusannahB.F.Paletz.2008}
\bibfield{author}{\bibinfo{person}{{Susannah B.F. Paletz}} {and} \bibinfo{person}{{Kaiping Peng}}.} \bibinfo{year}{2008}\natexlab{}.
\newblock \showarticletitle{Implicit theories of creativity across cultures: Novelty and appropriateness in two product domains}.
\newblock \bibinfo{journal}{\emph{Journal of Cross-Cultural Psychology}} \bibinfo{volume}{39}, \bibinfo{number}{3} (\bibinfo{year}{2008}), \bibinfo{pages}{286--302}.
\newblock
\showISSN{0022-0221}
\href{https://doi.org/10.1177/0022022108315112}{doi:\nolinkurl{10.1177/0022022108315112}}


\bibitem[Vaccaro et~al\mbox{.}(2024)]%
        {Vaccaro.2024}
\bibfield{author}{\bibinfo{person}{Michelle Vaccaro}, \bibinfo{person}{Abdullah Almaatouq}, {and} \bibinfo{person}{Thomas Malone}.} \bibinfo{year}{2024}\natexlab{}.
\newblock \showarticletitle{When combinations of humans and AI are useful: A systematic review and meta-analysis}.
\newblock \bibinfo{journal}{\emph{Nature Human Behaviour}} \bibinfo{volume}{8}, \bibinfo{number}{12} (\bibinfo{year}{2024}), \bibinfo{pages}{2293--2303}.
\newblock
\href{https://doi.org/10.1038/s41562-024-02024-1}{doi:\nolinkurl{10.1038/s41562-024-02024-1}}


\bibitem[Viechtbauer(2010)]%
        {Viechtbauer.2010b}
\bibfield{author}{\bibinfo{person}{Wolfgang Viechtbauer}.} \bibinfo{year}{2010}\natexlab{}.
\newblock \showarticletitle{Conducting meta-analyses in R with the metafor package}.
\newblock \bibinfo{journal}{\emph{Journal of Statistical Software}} \bibinfo{volume}{36}, \bibinfo{number}{3} (\bibinfo{year}{2010}), \bibinfo{pages}{1--48}.
\newblock
\href{https://doi.org/10.18637/jss.v036.i03}{doi:\nolinkurl{10.18637/jss.v036.i03}}


\bibitem[Viechtbauer and Cheung(2010)]%
        {Viechtbauer.2010}
\bibfield{author}{\bibinfo{person}{Wolfgang Viechtbauer} {and} \bibinfo{person}{Mike W.-L. Cheung}.} \bibinfo{year}{2010}\natexlab{}.
\newblock \showarticletitle{Outlier and influence diagnostics for meta-analysis}.
\newblock \bibinfo{journal}{\emph{Research Synthesis Methods}} \bibinfo{volume}{1}, \bibinfo{number}{2} (\bibinfo{year}{2010}), \bibinfo{pages}{112--125}.
\newblock
\showISSN{1759-2879}
\href{https://doi.org/10.1002/jrsm.11}{doi:\nolinkurl{10.1002/jrsm.11}}


\bibitem[Wadinambiarachchi et~al\mbox{.}(2024)]%
        {Wadinambiarachchi.2024}
\bibfield{author}{\bibinfo{person}{Samangi Wadinambiarachchi}, \bibinfo{person}{Ryan~M. Kelly}, \bibinfo{person}{Saumya Pareek}, \bibinfo{person}{Qiushi Zhou}, {and} \bibinfo{person}{Eduardo Velloso}.} \bibinfo{year}{2024}\natexlab{}.
\newblock \showarticletitle{The effects of generative AI on design fixation and divergent thinking}. In \bibinfo{booktitle}{\emph{CHI Conference on Human Factors in Computing Systems}} \emph{(\bibinfo{series}{CHI})}.
\newblock
\showISBNx{9798400703300}
\href{https://doi.org/10.1145/3613904.3642919}{doi:\nolinkurl{10.1145/3613904.3642919}}


\bibitem[Wang et~al\mbox{.}(2024)]%
        {Wang.132024}
\bibfield{author}{\bibinfo{person}{Haonan Wang}, \bibinfo{person}{James Zou}, \bibinfo{person}{Michael Mozer}, \bibinfo{person}{Anirudh Goyal}, \bibinfo{person}{Alex Lamb}, \bibinfo{person}{Linjun Zhang}, \bibinfo{person}{Weijie~J. Su}, \bibinfo{person}{Zhun Deng}, \bibinfo{person}{Michael~Qizhe Xie}, \bibinfo{person}{Hannah Brown}, {and} \bibinfo{person}{Kenji Kawaguchi}.} \bibinfo{year}{2024}\natexlab{}.
\newblock \bibinfo{title}{Can AI be as creative as humans?}
\newblock
\urldef\tempurl%
\url{https://arxiv.org/abs/2401.01623v4}
\showURL{%
\tempurl}


\bibitem[Wenger and Kenett(2025)]%
        {Wenger.1312025}
\bibfield{author}{\bibinfo{person}{Emily Wenger} {and} \bibinfo{person}{Yoed Kenett}.} \bibinfo{year}{2025}\natexlab{}.
\newblock \bibinfo{title}{We're different, we're the same: Creative homogeneity across LLMs}.
\newblock
\urldef\tempurl%
\url{https://arxiv.org/abs/2501.19361v1}
\showURL{%
\tempurl}


\bibitem[Wilson et~al\mbox{.}(1953)]%
        {Wilson.1953}
\bibfield{author}{\bibinfo{person}{Robert~C. Wilson}, \bibinfo{person}{Joy~P. Guilford}, {and} \bibinfo{person}{Paul~R. Christensen}.} \bibinfo{year}{1953}\natexlab{}.
\newblock \showarticletitle{The measurement of individual differences in originality}.
\newblock \bibinfo{journal}{\emph{Psychological Bulletin}} \bibinfo{volume}{50}, \bibinfo{number}{5} (\bibinfo{year}{1953}), \bibinfo{pages}{362--370}.
\newblock
\showISSN{0033-2909}
\href{https://doi.org/10.1037/h0060857}{doi:\nolinkurl{10.1037/h0060857}}


\bibitem[Wu et~al\mbox{.}(2025)]%
        {Wu.2025}
\bibfield{author}{\bibinfo{person}{Zhikun Wu}, \bibinfo{person}{Thomas Weber}, {and} \bibinfo{person}{Florian M{\"u}ller}.} \bibinfo{year}{2025}\natexlab{}.
\newblock \showarticletitle{One does not simply meme alone: Evaluating co-creativity between LLMs and humans in the generation of humor}. In \bibinfo{booktitle}{\emph{International Conference on Intelligent User Interfaces}} \emph{(\bibinfo{series}{IUI})}.
\newblock
\showISBNx{9798400713064}
\href{https://doi.org/10.1145/3708359.3712094}{doi:\nolinkurl{10.1145/3708359.3712094}}


\bibitem[Xu et~al\mbox{.}(2024)]%
        {Xu.662024}
\bibfield{author}{\bibinfo{person}{Cheng Xu}, \bibinfo{person}{Shuhao Guan}, \bibinfo{person}{Derek Greene}, {and} \bibinfo{person}{M-Tahar Kechadi}.} \bibinfo{year}{2024}\natexlab{}.
\newblock \bibinfo{title}{Benchmark data contamination of large language models: A survey}.
\newblock
\urldef\tempurl%
\url{https://arxiv.org/abs/2406.04244v1}
\showURL{%
\tempurl}


\bibitem[Zhang et~al\mbox{.}(2025)]%
        {Zhang.3302025}
\bibfield{author}{\bibinfo{person}{Man Zhang}, \bibinfo{person}{Ying Li}, \bibinfo{person}{Yang Peng}, \bibinfo{person}{Yijia Sun}, \bibinfo{person}{Wenxin Guo}, \bibinfo{person}{Huiqing Hu}, \bibinfo{person}{Shi Chen}, {and} \bibinfo{person}{Qingbai Zhao}.} \bibinfo{year}{2025}\natexlab{}.
\newblock \bibinfo{title}{AI delivers creative output but struggles with thinking processes}.
\newblock
\urldef\tempurl%
\url{https://arxiv.org/abs/2503.23327v1}
\showURL{%
\tempurl}


\bibitem[Zheng et~al\mbox{.}(2024)]%
        {Zheng.2024}
\bibfield{author}{\bibinfo{person}{Jiexin Zheng}, \bibinfo{person}{Ka~Chau Wong}, \bibinfo{person}{Jiali Zhou}, {and} \bibinfo{person}{Tat~Koon Koh}.} \bibinfo{year}{2024}\natexlab{}.
\newblock \showarticletitle{Large language model in ideation for product innovation: An exploratory comparative study}.
\newblock \bibinfo{journal}{\emph{Available at SSRN 4729982}} (\bibinfo{year}{2024}).
\newblock
\href{https://doi.org/10.2139/ssrn.4729982}{doi:\nolinkurl{10.2139/ssrn.4729982}}


\bibitem[Zhou and Lee(2024)]%
        {Zhou.2024}
\bibfield{author}{\bibinfo{person}{Eric Zhou} {and} \bibinfo{person}{Dokyun Lee}.} \bibinfo{year}{2024}\natexlab{}.
\newblock \showarticletitle{Generative artificial intelligence, human creativity, and art}.
\newblock \bibinfo{journal}{\emph{PNAS Nexus}} \bibinfo{volume}{3}, \bibinfo{number}{3} (\bibinfo{year}{2024}), \bibinfo{pages}{pgae052}.
\newblock
\href{https://doi.org/10.1093/pnasnexus/pgae052}{doi:\nolinkurl{10.1093/pnasnexus/pgae052}}


\bibitem[Zou and Zhu(2025)]%
        {Zou.2025}
\bibfield{author}{\bibinfo{person}{Wenbo Zou} {and} \bibinfo{person}{Feng Zhu}.} \bibinfo{year}{2025}\natexlab{}.
\newblock \showarticletitle{Generative AI adoption in human creative tasks: Experimental evidence}.
\newblock \bibinfo{journal}{\emph{Available at SSRN 5196748}} (\bibinfo{year}{2025}).
\newblock
\href{https://doi.org/10.2139/ssrn.5196748}{doi:\nolinkurl{10.2139/ssrn.5196748}}


\end{thebibliography}

\newpage
\textbf{Acknowledgment of AI Usage \& Overview of tools used}
During the preparation of this paper, generative AI tools---specifically OpenAI's ChatGPT---were used to assist with phrasing refinement, grammar editing, and the generation of illustrative code snippets. All intellectual contributions, including study design, analysis, interpretation of results, and the generation of core ideas, were made by the listed human authors. The AI was used solely as a supportive tool under human direction and supervision. In accordance with ACM's Policy on Authorship, the tool is not listed as an author, and its use is transparently disclosed here.\\[1.5em]

\textbf{Review Protocol.} No review protocol was preregistered for this study.\\[1.5em]

\textbf{Funding.} This research received no external funding. \\[1.5em]

\textbf{Competing Interests.} The authors declare no competing interests.\\[1.5em]

\textbf{Data Availability.} Code, data, and outputs are available via our Git at\\ \url{https://github.com/SM2982/Meta-Analysis-LLMs-Creativity.git}.
\end{document}